\documentclass[acmslarge,10.5pt]{acmart}

\AtBeginDocument{%
  }

\usepackage{verbatim}
\theoremstyle{definition}

\usepackage{amsthm}
\usepackage{threeparttable}
\usepackage{multirow}
\usepackage{tikz}
\usepackage{pgfplots}
\usetikzlibrary{plotmarks}
\usepackage{pgfplots}
\pgfplotsset{compat=1.12}
\usetikzlibrary{matrix}
\usepgfplotslibrary{groupplots}
\pgfplotsset{compat=newest}
\usepackage{subcaption}
\usepackage{graphicx}
\usepackage{longtable}
\usepackage{array}
\usepackage{hyperref}

\usepackage[ruled,linesnumbered]{algorithm2e}
\usepackage{boondox-cal}

\usepackage{tikz}
\usetikzlibrary{positioning}
\usepgfplotslibrary{statistics}

\begin{document}

\title{Leveraging Knowledge Graph Embedding for Effective Conversational Recommendation}

\author{Yunwen Xia}
\affiliation{
  \institution{School of Computer Science and Engineering, Nanyang Technological University}
  %\streetaddress{1 Th{\o}rv{\"a}ld Circle}
  %\city{Shanghai}
  \country{Singapore}}
\email{yunwen.xia@ntu.edu.sg}

\author{Hui Fang}
\authornote{Corresponding author}
\affiliation{
  \institution{RIIS \& SIME, Shanghai University of Finance and Economics}
  %\streetaddress{1 Th{\o}rv{\"a}ld Circle}
  \city{Shanghai}
  \country{China}}
\email{fang.hui@mail.shufe.edu.cn}

\author{Jie Zhang}
\affiliation{%
  \institution{School of Computer Science and Engineering, Nanyang Technological University}
  %\streetaddress{1 Th{\o}rv{\"a}ld Circle}
  %\city{Hekla}
  \country{Singapore}}
\email{zhangj@ntu.edu.sg}

\author{Chong Long}
\affiliation{%
  \institution{China Mobile Research Institute}
  \city{Beijing}
  \country{China}
}
\email{ longchong@chinamobile.com}

\renewcommand{\shortauthors}{Y. Xia et al.}

\begin{abstract}
%%%%%%%%%%%%%%%%%%modify%%%%%%%%%%%%%%%%%%%%%%%%%%%%%%%%%%%%%%%%%%%%%%%%%%%%%%%%%
Conversational recommender system (CRS), which combines the techniques of dialogue system and recommender system, has obtained increasing interest recently. In contrast to traditional recommender system, it learns the user preference better through interactions (i.e. conversations), and then further boosts the recommendation performance.
However, existing studies on CRS ignore to address the relationship among attributes, users, and items effectively, which might lead to inappropriate questions and inaccurate recommendations.
In this view, we propose a knowledge graph based conversational recommender system (referred as KG-CRS). Specifically, we first integrate the user-item graph and item-attribute graph into a dynamic graph, i.e., dynamically changing during the dialogue process by removing negative items or attributes. We then learn informative embedding of users, items, and attributes by also considering propagation through neighbors on the graph.
Extensive experiments on three real datasets validate the superiority of our method over the state-of-the-art approaches in terms of both the recommendation and conversation tasks.
\end{abstract}





\maketitle

\section{Introduction}
\label{sec:introdution}
Recommender systems have been widely applied in our daily life, including e-commerce, music and movies, restaurants, and online news, etc. Traditional recommender systems infer user’s preference based on user's interaction history or the history of similar users, which generally suffers from the cold-start problem and faces the challenge of dynamic user preference.
In this case,
\cite{christakopoulou2016towards} proposed conversational recommender system (CRS), which integrates the techniques of dialogue system and recommender system, and has attracted numerous research interest in recent years. That is, compared to traditional recommender systems, CRS learns user’s preference through several turns of communication (i.e., conversations), and then makes effective recommendations. Such a framework allows the platform (or recommendation agent) to make recommendations relatively easily for users without interaction history or users who dynamically change their preferences.
On the other hand, compared with task-oriented dialogue system \cite{dhingra2016towards} which helps users search for ideal items in a domain-related database/system, CRS can better understand users' needs based on interaction history and the history of similar users.

Overall, previous work on CRS can mainly be classified into two categories. One category \cite{chen2019towards, li2018towards, 48414, liao2019deep, zhou2020improving} is about Natural Language Processing (NLP) in CRS, which attempts to
understand users’ utterances and merge each recommended item into a textual sentence generated
by a CRS system. Thus, it allows users to express their questions more freely and strives to make
the system’s responses more like human languages. Conversely, the other category \cite{sun2018conversational, lei2020estimation, lei2020interactive, christakopoulou2016towards, xu2020user, zhang2018towards, zou2020towards} focuses on ``strategy'' in CRS. Here, ``strategy'' refers to what action to conduct in each turn (conversation), such as ``whether to recommend'', ``which items to recommend'', or ``which attributes to ask''.  
Most studies in this category choose to simplify (e.g., using template-based sentences) or just skip the NLP module. Our study falls into this category and considers that it is feasible to add the NLP module derived from the first category of studies into our CRS system if needed. 

The dialogue ``strategy'' has not been well discussed among the first category of studies, but it is of great importance for CRS. An effective ``strategy''
is expected to let CRS perform well in both recommendation task and conversation task, i.e., recommend satisfactory items to users while avoid asking useless questions about attributes to infer users' preferences (i.e., maintain fewer rounds of conversations). 

As it is difficult to define the best ``strategy'' in CRS, reinforcement learning (RL) is adopted in the framework \cite{sun2018conversational, lei2020estimation, lei2020interactive}. One of the key issues in these methods is how to represent users, items, and attributes, as these representations are the basis of both recommendation and ``strategy'', and will directly affect the results of recommendation and conversation tasks.
Take user representation as an example, a user is profiled by his/her interaction history with items, thus learning the corresponding user representation should also be entangled with the interaction history. Previous work \cite{sun2018conversational, lei2020estimation, lei2020interactive} mainly uses user representation to predict users' possible future interactions and argued that the learned user representation might \emph{implicitly} capture the historical interactions. However, such kind of implicit modeling might lead to difficulty in well capturing the relationships among users, items and attributes, and the learning process would be further deteriorated by the data imbalance issue. That is, the interaction intensities among different users, items, and attributes are quite dissimilar.
Therefore, we argue that it is worthwhile to explicitly encode the interaction history to provide better guidance and explanations for representation learning. 

In this view, we propose a \underline{K}nowledge \underline{G}raph based \underline{C}onversational \underline{R}ecommender \underline{S}ystem (KG-CRS), which adopts knowledge graph embedding techniques to explicitly encode the connections among users, items and attributes for more accurate representations, effective recommendations, and efficient conversations.
In particular, we combine the user-item graph (identified from users' historical interactions) and item-attribute graph together, then learn user representations, item representations and attribute representations by also considering the propagation according to the graph. 
This provides guidance and explanations for the learned representations and greatly facilitates the representation learning of users, items and attributes, which are the basis of the recommendation and conversation tasks.

Besides, to further enhance the conversation process, we dynamically update the whole graph by removing negative items or attributes identified by the previous conversations.
Extensive experiments on three real-world datasets validate the superiority of our method over the state-of-the-art approaches in terms of both the recommendation and conversation tasks.

The contribution of our work is threefold:
\begin{itemize}
\item We contribute to the recommendation task in CRS by utilizing graph embedding techniques to better learn user and item representations, and thus greatly boost the recommendation performance in terms of accuracy metrics.

\item We contribute to the conversation task in CRS by, besides the effective attribute representations via graph embedding, also exploiting the information of similar users for asking more effective questions. Besides, we consider a dynamic graph framework by removing negative items and attributes to avoid asking useless questions.

\item 
In our experiments, we firstly consider a new metric, average positive action (APA), to measure user satisfaction in the conversation process. 
In initializing the parameters of our policy network, besides the traditional max-entropy mimic strategy, we propose to also consider the ground-truth mimic strategy. Our extensive experiments can serve as guidance for the following research on CRS.
\end{itemize}

The rest of the paper is organized as follows. We first review the literature on conversational recommender system and KG-based recommendation system in Section \ref{sec:relatedwork}. We then formulate our research problem and present the details of our KG-CRS framework in Section \ref{sec:model}. After that, we conduct extensive experiments on three real datasets to verify the effectiveness of our approach in Section \ref{sec:experiments}. Finally, we conclude our paper in Section \ref{sec:conclusions}.

\section{Related Work}
\label{sec:relatedwork}

In this section, we mainly discuss two folds of related work: (1) conversational recommender system; and (2) knowledge graph (KG)-based recommender system.
\subsection{Conversational Recommender System}
\label{sec:crsliterature}
As introduced before, the existing studies on CRS can be concluded into two categories, since the work of ~\cite{christakopoulou2016towards}. 

% Another line of research emphasizes 
The first line focuses on the understanding and generation of conversations from the perspectives of NLP and considers the semantic information in CRS \cite{chen2019towards}.
Specifically, it attempts to understand users' utterances and merge recommended items into textual sentences generated by CRS. Thus, users are allowed to express their questions more freely, meanwhile the system's responses are more in a human language form. 
For example, Li et al. \cite{li2018towards} and Radlinski et al. \cite{48414} collected datasets consisting of real-world conversations to facilitate the NLP research on CRS.
The work of \cite{liao2019deep} considered the item-attribute graph for multiple recommendation tasks (e.g., both hotel and restaurant recommendations), while Zhou et al. \cite{zhou2020improving} semantically fused both item-oriented graph and word-oriented graph for CRS. 
Yu et al. \cite{yu2019visual} further combined NLP with visual information.

There are two main drawbacks with existing studies of this category: (1) the training process relies heavily on the manually collected natural language data, which might hinder its usability to a new domain where new data should be collected to retrain the corresponding methods from scratch.
%What's more, as the training process of these methods is to let the model mimic the natural language data, they 
Besides, it ignores to well explore the ``strategy'' in CRS;
%If the conversation data we collected is not reasonable enough, such as making a recommendation too late, the model may also learn such a bad ''strategy''. 
and (2) user's interaction history is not well utilized in the previous studies in this category. As a result, a CRS system may generate similar conversations for different users, which is thus not personalized.

Another line of research emphasizes the effective ``strategy'' in CRS, which includes four main sub-tasks: ``when to recommend'', ``which items to recommend'', ``when to ask'', and ``which attributes to ask''.
For example, Zou et al. \cite{zou2020towards} improved the solution on ``which items to recommend'' via a novel matrix factorization model, and the task ``which attributes to ask'' was addressed through Generalized Binary Search (GBS). For the other two sub-tasks, it chooses to keep asking until the size of candidate items is smaller than the size of recommendation list or the maximum number of questions is reached. 
Such ``strategy'' is similar to Max Entropy in \cite{sun2018conversational} and the method in \cite{zou2018technology, zou2019learning, zou2020towardsTAR, bi2019conversational}, the main difference is the recommendation module. Bi et al. \cite{bi2019conversational} utilized the negative feedback during the conversation to improve the performance of recommendation task. Zou et al. \cite{zou2018technology, zou2019learning, zou2020towardsTAR} ranked items based on Bayes' rule. 
However, most of these methods are not flexible enough for ``when to recommend'' problem, and fail to consider that some users might be made successful recommendations in the first few rounds of conversations. Besides, it is not personalized to address ``which attributes to ask'' with heuristic methods (e.g. GBS, Max Entropy). 

In addition to these studies on ``which items to recommend'', several studies \cite{aliannejadi2019asking, krasakis2020analysing} pay attention to ``which attributes to ask''.
However, we argue that the four sub-tasks should be discussed together, as the conversation task and the recommendation task affect and interact with each other. Specifically, if the recommendation has been successfully made in the first few rounds of conversations, then the subsequent conversations need not be carried out. 
In the meantime, more effective and accurate conversations would improve recommendation performance.
Zhang et al. \cite{zhang2018towards} encoded items based on their textual descriptions and implemented a conversational search system from the perspective of information retrieval. Xu et al. \cite{xu2020user} constructed a personalized user memory graph for conversational recommendation. 
However, these studies did not explore the ``strategy'' for the four tasks and inclined to only follow the behaviour in terms of the real datasets. In this case, the performance may deteriorate when the behaviour in the datasets is unexpected and relatively unreasonable.

On the other hand, several studies \cite{sun2018conversational, lei2020estimation, lei2020interactive} use reinforcement learning to train a policy network for making decisions on the four tasks in CRS.
For example, Sun and Zhang \cite{sun2018conversational} considered a single-round recommendation scenario, where once a recommendation is made, the dialogue will end regardless of whether the recommendation is successful or not.
On the contrary,
Lei et al. \cite{lei2020estimation, lei2020interactive} considered a more realistic setting, i.e., multi-round recommendation scenario, which means the dialogue will terminate only when the recommendation is successful or the predefined termination condition (i.e., the maximum number of rounds) happens.
In \cite{lei2020interactive}, it further considers the graph-based conversational path reasoning to help narrow the candidate set of attributes and items.
The aforementioned three studies \cite{sun2018conversational, lei2020estimation, lei2020interactive} are the most relevant to our work, as we also consider the reinforcement learning framework and multi-round recommendation scenario for the conversation and recommendation tasks.
We consider that these methods \cite{sun2018conversational, lei2020estimation, lei2020interactive} can still be improved from more accurate entity representations (e.g., users, items and attributes). 
The corresponding representation learning in \cite{sun2018conversational, lei2020estimation, lei2020interactive} does not make full use of the relations among users, items and attributes. 
For example, during the learning process of user representation, they \cite{sun2018conversational, lei2020estimation, lei2020interactive} utilize users' interaction history with items as the model's final target, with the expectation that the learned representations can implicitly capture the relations between users and items. However, such kind of implicit
modeling might lead to difficulty in well capturing the relationships between users and items, and the learning process would be further deteriorated by the data imbalance issue.
It should be noted that although Lei et al. \cite{lei2020interactive} introduced a knowledge graph for CRS, the graph is merely used to narrow the candidate set of items or attributes instead of being involved to learn better representations.

\subsection{KG-based Recommender System}
Traditional recommender system has been widely studied. These studies might be helpful for building an effective CRS, where the recommendation task is of great importance. We only explore the KG-based ones as they are most relevant to our study. Those studies that utilize KG to improve the performance of CRS \cite{liao2019deep, zhou2020improving, xu2020user, lei2020interactive} have been introduced in Section \ref{sec:crsliterature}, thus, their recommendation module will not be repeatedly discussed here. We merely review the KG-related traditional recommender systems here.

Some studies \cite{zhang2016collaborative, wang2018dkn, huang2018improving} directly utilize mature Knowledge Graph Embedding (KGE) methods to integrate side information for recommendation. For example, Zhang et al.\cite{zhang2016collaborative} combined the structural data (graph), textual data and visual data to improve the item representations, where the structural data is explored using TransR \cite{lin2015learning}. Wang et al. \cite{wang2018dkn} also applied translation-based methods (e.g., TransR \cite{lin2015learning}, TransD \cite{ji2015knowledge}) to learn the embedding of knowledge entities in news titles, and then designed a Knowledge-aware CNN for generating recommendations on news. Similarly, Huang et al. \cite{huang2018improving} incorporated knowledge graph information via TransE \cite{bordes2013translating} to improve sequential recommendation.
The aforementioned methods, directly applying KGE, are validated to be effective in integrating side information for recommendation, but they ignore to well capture the interactions between users and items in graph learning.

In contrast, path-based methods \cite{yu2014personalized, zhao2017meta, sun2018recurrent, hu2018leveraging} better utilize the relationship among entities in the graphs. For example, the studies of \cite{yu2014personalized,zhao2017meta} design various meta-paths (e.g., $user \rightarrow movie \rightarrow actor \rightarrow movie$) in the graphs to estimate the probability between every user and every item.
Sun et al. \cite{sun2018recurrent} also extracted various meta-paths between a user and an item, and then integrated different path through a recurrent neural network and a pooling layer. A fully-connected layer is then applied to calculate the recommendation score between the user and the item. 
Furthermore, Hu et al. \cite{hu2018leveraging} learned explicit meta-path representations with co-attention mechanism to enhance the recommendation task.

As the path-based methods require the manually calibrated design of the meta-paths, which might be relatively difficult in some scenarios, the propagation-based methods \cite{wang2018ripplenet, wang2019kgat} have thus been proposed. For example, Wang et al. \cite{wang2018ripplenet} took some nodes (i.e., items historically clicked by a user) as the center, explored high-order information in the graph, and finally integrated the exploration result of each layer as the user representation. Similarly, Wang et al. \cite{wang2019kgat} iteratively transferred the neighbourhood information to a node to update the embedding of this node, and combined the result after each update as the representation of that node. After learning the representation of each user and each item, both \cite{wang2018ripplenet} and \cite{wang2019kgat} adopt the inner product to estimate the recommendation scores.

The main difference between our work and traditional KG-based recommender system is that, we not only model the relationship between each candidate item and a user, but also address the relationship between each candidate item and the positive attributes confirmed by the user.

\section{Proposed Model}
\label{sec:model}
In this section, we present our KG-CRS model in great detail by first simply formulating the multi-round recommendation scenario.

\subsection{Problem Formulation}
\begin{figure}[t]
\centering
\includegraphics[scale=0.38]{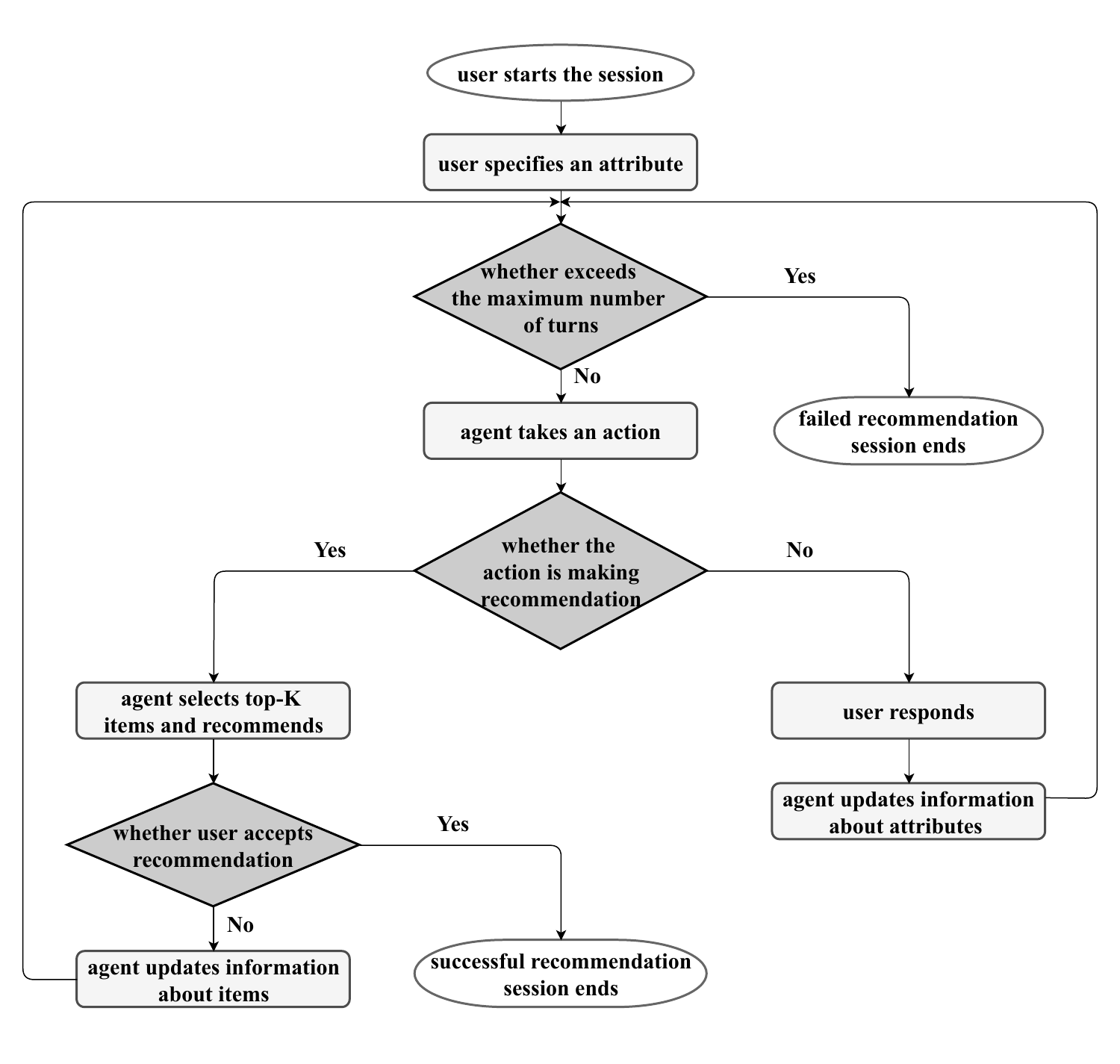}
\caption{A session of Conversational Recommender System.}
\label{fig:workflow}
\end{figure}

The same as \cite{lei2020estimation}, we focus on the multi-round recommendation scenario where a CRS interacts with a user multiple times (together arranged as a \emph{session}) by either asking attributes or recommending items until the user is satisfied with the recommendation or chooses to leave the session without a satisfactory recommendation. It should be noted that, similar to \cite{zhang2018towards, sun2018conversational, lei2020estimation, lei2020interactive}, the conversation process adopts ``System ask, User Respond'' pattern. That is to say, a conversation session is mostly driven by the system, except for the beginning and the end of the session. More detailed process is shown in Figure \ref{fig:workflow}, and the main notations are summarized in Table \ref{tb:notations}.

Specifically, at the beginning of each session, user $u$ firstly specifies a favorable \emph{attribute} $p$ (from the set of attributes $P$ describing items on the platform , $p\in P$), and the CRS shortlists the candidate items containing attribute $V_p\subset V$. Next, in each round (turn), the CRS conducts an \emph{action} $a \in A$: \emph{asking attribute} or \emph{making recommendation}. That is, for the action of making recommendation, if the user accepts the recommendation, the session will be terminated at the corresponding round. Otherwise, the recommended items are considered as \emph{negative items}, and we move to the next round. On the other hand, for the action of asking attribute, if the user considers the asked attribute in this round is irrelevant to her preference, the attribute will be marked as \emph{negative attribute} and we also move to the next round, otherwise the attribute is considered as \emph{positive attribute}. The action space $|A|$ is equal to ${N_{q}+1}$, where ${N_{q}}$ represents the number of questions about attributes. That is to say, when the system decides to ask attributes, it actually selects a question in the question pool of size ${N_{q}}$ and then asks. To avoid repetitive questions, those questions that have been asked will be recorded and eliminated during the conversation session.

\begin{table}[t]
\centering
\small
  \caption{The main notations.}
  \label{tb:notations}
  \begin{tabular}{m{2cm}<{\centering}|m{7cm}<{\raggedright}}
%\begin{tabular}{c|l}
    \hline
    Notations& Explanation\\
    \hline
  
    $U$, $V$, $P$ & the set of all users, the set of all items, the set of all attributes \\
    $u$, $v$, $p$ & user $u\in U$, item $v\in V$, attribute $p\in P$ \\
    $P_u$ & the set of attributes confirmed by user in a session \\
    $P_{neg}$ & the set of negative attributes that user does not care in the session \\
    $P_{ses}$ & the preference of user $u$ in the session \\
    $V_p$ & the set of items that contains the attribute $p$ \\
    $V_{cand}$ & the set of candidate items during the session \\
    $V_{neg}$ & the set of negative items during the session \\
    $V_{rec}$ & the set of items to be recommended \\
 %   $t$ & turn number \\
    $T$, $T'$ & the maximum number of turns for the session, actual number of turns in the session \\
    $K$ & the size of items to be recommended in each turn \\
    $N_q$ & the number of questions about attributes that can be asked by the CRS agent\\
    $A$ & the set of all actions that the CRS agent can take, $|A|=N_q+1$ \\
    $a_t$ & action at turn $t$ \\
    $a_p$ & action about attribute $p$ \\
    $P_{a}$ & the set of attributes that are related to action $a$, $a\in A$ \\
    $A_{ask}$ & the set of actions which represents asking attributes, and its size is $N_q$ \\
    $A_{know}$ & the set of actions which have been taken during the session except for the recommendation\\
    $A_{cand}$ & the set of candidate actions in the session\\
    $\mathcal{G}$ & knowledge graph $\mathcal{G}$\\
    $\mathcal{h}$, $\mathcal{t}$, $\mathcal{r}$ & head entities, tail entities and relations in the graph \\
    $\mathcal{N_h}$ & neighbourhood of entity $\mathcal{h}$ \\
    $\alpha_{\mathcal{h},\mathcal{r},\mathcal{t}}$ & attention weight of different tail entities $\mathcal{t}$ for head entity $\mathcal{h}$ \\
    $\mathcal{K}$ & propagation step in the graph \\
    $e_{\mathcal{h}}$, $e_{\mathcal{r}}$, $e_{\mathcal{t}}$ & embedding of entity $\mathcal{h}$, relation $\mathcal{r}$ and entity $\mathcal{t}$ \\
    $e_{\mathcal{h}}{'}$, $e_{\mathcal{t}}{'}$ & embedding of entity $\mathcal{h}$ and entity $\mathcal{t}$ after propagation \\
    $M_{\mathcal{rh}}$, $M_{\mathcal{rt}}$, $m_{\mathcal{h}}$, $m_{\mathcal{t}}$, $m_{\mathcal{r}}$ & the projection matrices and the projection vectors to determine corresponding projection matrices \\
    $E_{\mathcal{h},\mathcal{r},\mathcal{t}}$ & parameters of TransD model, including the projection matrices and the embedding \\ 
    $W$, $b$ & parameters to integrate the information of entities and its neighbourhood\\
    $\theta$ & parameters of the policy network \\
    $loc_{t}$ & the location of ideal item at turn $t$ \\
    $y_{v}$, $y_{p}$ & score of item $v$ and attribute $p$ \\
    $L_{graph}$, $L_{item}$, $L_{conv}$ & loss function \\
    $r_t$, $r_{attr}$, $r_{item}$, $r_{turn}$, $r_{quit}$, $r_{imp}$ & reward at turn $t$ and its five components \\
    $\textbf{s}$, $\textbf{s}_{ent}$, $\textbf{s}_{user}$, $\textbf{s}_{conv}$, $\textbf{s}_{dial}$ & state vector and its four components \\
    $\textbf{h}$, $\textbf{u}$, $\textbf{v}$, $\textbf{p}$ & the embedding of entity $\mathcal{h}$ (user $u$, item $v$, attribute $p$) after propagation\\
  \hline
\end{tabular}
\end{table}

Given the multi-round scenario, the objective of our CRS is to provide satisfactory recommendations to each user in fewer rounds for each session. To fulfill the goal, we propose our KG-CRS framework, as elaborated below.

\subsection{The KG-CRS Framework}

\begin{figure*}[t]
\centering
\includegraphics[scale=0.35]{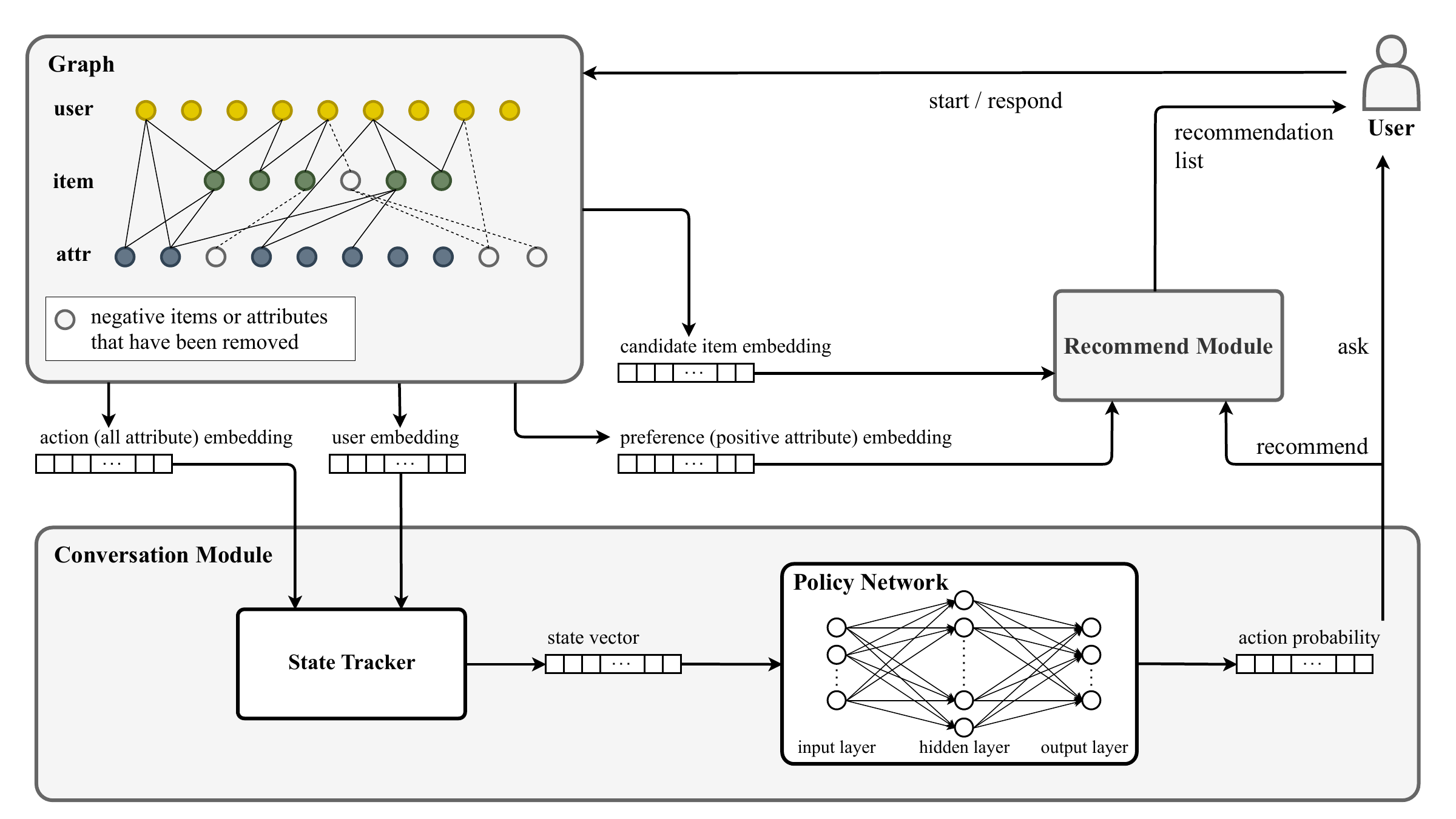}
\caption{The KG-CRS framework.}
\label{fig:kgFramework}
\end{figure*}

% Fig.1 Model Structure
The framework of our KG-CRS is demonstrated in Figure \ref{fig:kgFramework}. Specifically, it consists of a dynamic knowledge graph and two modules, a conversation module and a recommendation module.
The dynamic graph consists of three types of entities (namely user, item, and attribute) and three types of corresponding relations (denoted as user-item, item-attribute, and user-attribute). Its content will change (i.e. removing \emph{negative items} and \emph{negative attributes}) during a conversation session and will be reset when a new session starts.
The conversation module encodes the current \emph{state} and decides an \emph{action} (asking attribute or making recommendation) according to the current state, while the recommendation module will give recommendations from the candidate items when the conversation module decides to make recommendation. Both conversation and recommendation modules rely on the informative embedding learned from the dynamic graph, i.e., to encode the current state in conversation module, and to calculate the recommendation score of each candidate item.
In this framework, a conversation session is both started and ended by a user, and the user will give responses when the CRS agent asks questions. 
Noted that, similar to \cite{lei2020estimation, lei2020interactive}, we skip the Natural Language Understanding (NLU) and Natural Language Generation (NLG) in our model.

In the followings, we introduce the graph embedding learning, and the two modules, namely, conversation module and recommendation module.

\subsubsection{Graph Embedding Learning.}

\begin{figure}[t]
\centering
\includegraphics[scale=0.2]{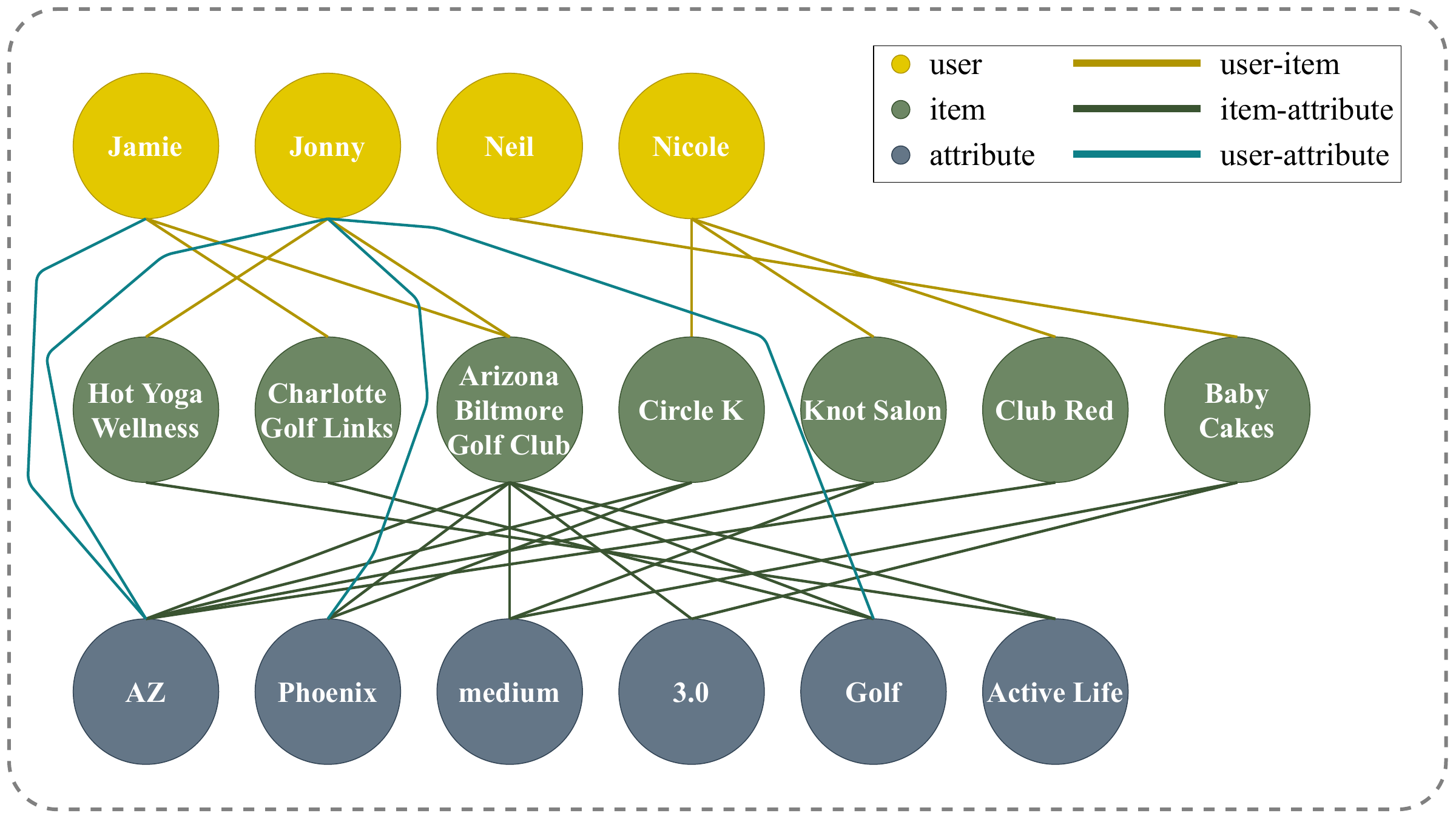}
\caption{An example of knowledge graph in our study.}
\label{fig:kg}
\end{figure}

As shown in Figure \ref{fig:kg}, there are three types of entities (namely user, item, and attribute) and three types of corresponding relations (denoted as user-item, item-attribute and user-attribute) in the graph $\mathcal{G} =\{ (\mathcal{h},\mathcal{r},\mathcal{t}) | \mathcal{h},\mathcal{t} \in \mathcal{E}, \mathcal{r} \in \mathcal{R} \}$, where $\mathcal{h}, \mathcal{t}$ represent entities (head and tail respectively), and $\mathcal{r}$ represents relations. Every (entity, relation, entity) triple means that there is a relationship $\mathcal{r}$ between entity $\mathcal{h}$ (e.g., user) and entity $\mathcal{t}$ (e.g., item), for example, user `Neil' was interested in item `Baby Cakes'. Inspired by \cite{he2017learning, wang2019kgat, wang2019you}, the learning process of embedding is divided into two stages, node feature and information propagation. The design of 
dynamic graph is to prevent \emph{negative items} and \emph{negative attributes} from affecting other related entities' representation during the information propagation process.

\medskip\textbf{Node Feature.} This stage is to learn the vector representation of each entity and relation in the graph. We use a translation-based method, TransD \cite{ji2015knowledge, han2018openke}, as it performs better on describing the structure of knowledge graph. TransD assumes that $e_{\mathcal{h}}{'} + e_{\mathcal{r}} \approx e_{\mathcal{t}}{'}$ if triple $(\mathcal{h},\mathcal{r},\mathcal{t})$ exists in the graph, where $e_{\mathcal{h}}{'}, e_{\mathcal{t}}{'}$ are the embedding of entity $\mathcal{h}, \mathcal{t}$, and $ e_{\mathcal{r}}$ is the embedding of relation $\mathcal{r}$. Such assumption is achieved by optimizing the score function:
\begin{equation} \label{eq:transd_score}
    f(\mathcal{h}, \mathcal{r}, \mathcal{t}) = \|e_{\mathcal{h}}{'} + e_{\mathcal{r}} - e_{\mathcal{t}}{'} \|^2_2
\end{equation}
where $e_{\mathcal{h}}{'} = M_{\mathcal{rh}}e_{\mathcal{h}}$ and $e_{\mathcal{t}}{'} = M_{\mathcal{rt}}e_{\mathcal{t}}$. $M_{\mathcal{rh}}$ and $M_{\mathcal{rt}}$, as defined in Equation \ref{eq:projection_matrix}, are the projection matrices, which assure that each type of entity have different representations for different relations. $e_{\mathcal{h}}, e_{\mathcal{t}}$ are the embedding of entity $\mathcal{h}, \mathcal{t}$ without projection, respectively. The score $f(\mathcal{h}, \mathcal{r}, \mathcal{t})$ is low if there is a triple $(\mathcal{h},\mathcal{r},\mathcal{t})$, and high otherwise.
\begin{equation} \label{eq:projection_matrix}
    M_{\mathcal{rh}} = m_{\mathcal{r}}m_{\mathcal{h}}^T + I;\ M_{\mathcal{rt}} = m_{\mathcal{r}}m_{\mathcal{t}}^T + I
\end{equation}
where $m_{\mathcal{h}}$, $m_{\mathcal{t}}$ and $m_{\mathcal{r}}$ represent the projection vectors to determine corresponding projection matrices, and $I$ is the identity matrix to initialize the projection matrices.

The loss function of this stage is defined as:
\begin{equation} \label{eq:loss_graph}
    L_{graph} = \sum\max(f(\mathcal{h}, \mathcal{r}, \mathcal{t}) - f(\mathcal{h}, \mathcal{r}, \mathcal{t}{'}), -\lambda)
\end{equation}
where $\lambda$ is the margin parameter. Triple $(\mathcal{h},\mathcal{r},\mathcal{t})$ is the positive triple that exists in the graph, while triple $(\mathcal{h},\mathcal{r},\mathcal{t}{'})$ is a negative triple generated by randomly sampling an irrelevant entity $\mathcal{t}{'}$ for $\mathcal{h},\mathcal{r}$.

\medskip\textbf{Information Propagation.} This stage is to enrich the entity embedding by recursively updating based on each entity's neighbourhood. As illustrated in Figure \ref{fig:kg}, user `Jamie' may be interested in item `Hot Yoga Wellness' because both item `Arizona Biltmore Golf Club' and item `Hot Yoga Wellness' have attribute `Active Life'. Therefore, it is necessary to propagate the information about item `Arizona Biltmore Golf Club' to the user `Jamie' while learning her embedding. The neighbourhood $\mathcal{N_h}$ of entity $\mathcal{h}$ is defined as:
\begin{equation} \label{neighbour_embedding}
    \mathcal{N_h}=\sum_{(\mathcal{h},\mathcal{r},\mathcal{t}) \in \mathcal{G}}\alpha_{\mathcal{h},\mathcal{r},\mathcal{t}}e_{\mathcal{t}}
\end{equation}
where $e_{\mathcal{t}}$ is the embedding of tail entity $\mathcal{t}$ in the triple $(\mathcal{h},\mathcal{r},\mathcal{t})$ without projection. $\alpha_{\mathcal{h},\mathcal{r},\mathcal{t}}$ is the attention weight, which is computed according to the distance between $\mathcal{h}$ and $\mathcal{t}$:
\begin{equation} \label{neighbour_attention}
    \alpha_{\mathcal{h},\mathcal{r},\mathcal{t}} = e_{\mathcal{t}}{'}\tanh(e_{\mathcal{h}}{'} + e_{\mathcal{r}})^\top
\end{equation}

We recursively update the embedding of entity $\mathcal{h}$ as:
\begin{equation} \label{h_embedding_each_step}
   % \begin{align}%\label{h_embedding_each_step}
        e_{\mathcal{h}}^{(k)} = g(e_{\mathcal{h}}^{(k-1)}, \mathcal{N_h}^{(k-1)})
        = {\rm ReLU}(W^{k}(e_{\mathcal{h}}^{(k-1)} + \mathcal{N_h}^{(k-1)}) + b^{k})
  % \end{align}
\end{equation}
where $W^{k}$ is the parameter matrix, $b^{k}$ is a bias and $k=1, 2, ..., \mathcal{K}$. $e_{\mathcal{h}}^{(k-1)}$ and $\mathcal{N_h}^{(k-1)}$ represent the embedding of entity $\mathcal{h}$ and the information of neighbourhood $\mathcal{N_h}$ at step $k-1$, respectively. $e_{\mathcal{h}}^{(1)}=e_{\mathcal{h}}$, $\mathcal{N_h}^{(1)}=\mathcal{N_h}$.

The final embedding of entity $\mathcal{h}$ (Equation \ref{eq:final embedding}) is the concatenation of $e_{\mathcal{h}}^{(k)}$ at each step. In the following subsections, we will use $\textbf{u}$, $\textbf{v}$ and $\textbf{p}$ to represent the final embedding of the three different types of entities (user, item and attribute) in the graph. And, the calculation of the three representations all refers to Equation \ref{eq:final embedding}.
\begin{equation} \label{eq:final embedding}
   \textbf{h} = [e_{\mathcal{h}}^{(1)}, e_{\mathcal{h}}^{(2)}, ..., e_{\mathcal{h}}^{(\mathcal{K})}]
\end{equation}

\subsubsection{Recommendation Module.}
The main objective of this module is to obtain recommendation scores of candidate items ($V_{cand}$) based on the graph embedding, and simultaneously measure the loss function for graph embedding learning during the training process.
Specifically, it generates a score of each candidate item, sorts the items in descending order in terms of the scores, and finally recommends the top $K$ items to the target user. It should be noted that \emph{negative items} and \emph{negative attributes} will not be utilized explicitly when calculating the score. The removal of these negative entities will only change the embedding obtained in the information propagation stage. The score of each item $v \in V_{cand}$ is calculated as Equation \ref{eq:item_score}:
\begin{equation} \label{eq:item_score}
    y_v = \textbf{u}_{0}^T\textbf{v}_{0} + \sum_{p \in P_u} \textbf{v}_{1}^T\textbf{p}_{1}
    % y_v = \mathbf{u}^{ui}\mathbf{i}^{ui} + \sum_{a_{pos}\in A_{pos}}\mathbf{i}^{ia}\mathbf{a}_{pos}^{ia}
\end{equation}
where $\textbf{u}_{0}= M_{\mathcal{0u}}\textbf{u}$, $\textbf{v}_{0}= M_{\mathcal{0v}}\textbf{v}$, $\textbf{v}_{1}= M_{\mathcal{1v}}\textbf{v}$, and $\textbf{p}_{1}= M_{\mathcal{1p}}\textbf{p}$. Specifically, (1) \textbf{u}, \textbf{v} and \textbf{p} are the embedding of user $u$, item $v$ and positive attribute $p$, respectively; (2) $M_{\mathcal{0u}}$, $ M_{\mathcal{0v}}$, $ M_{\mathcal{1v}}$ and $\ M_{\mathcal{1p}}$ are the projection matrices in Equation \ref{eq:projection_matrix} to assure item $v$ to be capable of having different representations for different relations, namely, user-item (i.e., $0$ in the subscript, and $\mathcal{r}_0$ in Algorithm \ref{alg:embedding learning}) and item-attribute (i.e., $1$ in the subscript, and $\mathcal{r}_1$ in Algorithm \ref{alg:embedding learning}\footnote{$\mathcal{r}_2$ in Algorithm \ref{alg:embedding learning} refers to user-attribute relation.}), respectively.

The loss function in this module is defined as:
\begin{equation}
    L_{item} = \sum \log \sigma (y_v) - \log \sigma (y_v')
\end{equation}
where $y_v$ is the recommendation score of positive item $v$ that user $u$ has interacted with in the history, and $y_v'$ is the recommendation score of negative item $v'$.

The training process of graph embedding learning and recommendation module is detailed in Algorithm \ref{alg:embedding learning}. As elaborated, 
graph embedding is first pre-trained with $L_{graph}$ (line 1), and then optimized with $L_{graph}$ and $L_{item}$ alternately (lines 2-19). It should be noted that the graph embedding and other aforementioned parameters are learned in advance, and then kept fixed during the training process of the conversation module. They are only updated while there occurs negative items during the conversation session (see line 25 in Algorithm \ref{alg:conversation module}).

\begin{algorithm}
\caption{Offline Training of Graph Embedding and Recommendation Module}
\label{alg:embedding learning}
\SetKwInOut{Input}{Input}
\Input{the set of all embedding $E_{\mathcal{h},\mathcal{r},\mathcal{t}}$, propagation step $\mathcal{K}$, the set of parameters $W$ and $b$}
pretrain graph embedding $E_{\mathcal{h},\mathcal{r},\mathcal{t}}$ with $L_{graph}$ \;
update attention weight $\alpha_{\mathcal{h},\mathcal{r},\mathcal{t}}$ \;
\For{epoch = 1, 2, ..., $N_{epoch}$}{
    \For{batch = 1, 2, ..., $N_{batch}$}{
        load training data $(\mathcal{h},\mathcal{r},\mathcal{t})$, including $(u,\mathcal{r}_{0},v)$, $(u,\mathcal{r}_{0},v')$, $(v,\mathcal{r}_{1},p)$, $(v',\mathcal{r}_{1},p)$,
        $(u,\mathcal{r}_{2},p)$, $(u,\mathcal{r}_{2},p')$ \;
        calculate positive score $f(\mathcal{h}, \mathcal{r}, \mathcal{t})$ and negative score $f(\mathcal{h}, \mathcal{r}, \mathcal{t}{'})$ using Equation \ref{eq:transd_score}\;
        calculate loss function: $L_{graph} = \sum\max(f(\mathcal{h}, \mathcal{r}, \mathcal{t}) - f(\mathcal{h}, \mathcal{r}, \mathcal{t}{'}), -\lambda)$ \;
        \textbf{update} graph embedding $E_{\mathcal{h},\mathcal{r},\mathcal{t}}$ with $L_{graph}$ \;
        $e_{\mathcal{h}}^{(1)} = e_{\mathcal{h}}$, $e_{\mathcal{h}} \in E_{\mathcal{h},\mathcal{r},\mathcal{t}}$; $\mathcal{N_h}^{(1)}=\mathcal{N_h}=\sum_{(\mathcal{h},\mathcal{r},\mathcal{t}) \in \mathcal{G}}\alpha_{\mathcal{h},\mathcal{r},\mathcal{t}}e_{\mathcal{t}}$, $e_{\mathcal{t}} \in E_{\mathcal{h},\mathcal{r},\mathcal{t}}$ \;
        \For{k = 2, ..., $\mathcal{K}$}{
            $e_{\mathcal{h}}^{(k)} = g(e_{\mathcal{h}}^{(k-1)}, \mathcal{N_h}^{(k-1)}) = {\rm ReLU}(W^{k}(e_{\mathcal{h}}^{(k-1)} + \mathcal{N_h}^{(k-1)}) + b^{k})$; $W^{k} \in W$, $b^{k} \in b$ \;
            %\tcp*{Information Propagation}
            $\mathcal{N_h}^{(k)}=\sum_{(\mathcal{h},\mathcal{r},\mathcal{t}) \in \mathcal{G}}\alpha_{\mathcal{h},\mathcal{r},\mathcal{t}}e_{\mathcal{t}}^{(k)}$;         
        } 
        concatenate the result of propagation, $\textbf{h} = [e_{\mathcal{h}}^{(1)}, e_{\mathcal{h}}^{(2)}, ..., e_{\mathcal{h}}^{(\mathcal{K})}]$ \;
        calculate positive score $y_v$ and negative score $y_v'$ as Equation \ref{eq:item_score} \;
        calculate loss function: $L_{item} = \sum \log \sigma (y_v) - \log \sigma (y_v')$\;
        \textbf{update} graph embedding $E_{\mathcal{h},\mathcal{r},\mathcal{t}}$, parameters $W$ and $b$ with $L_{item}$ \;
    }
    update attention weight $\alpha_{\mathcal{h},\mathcal{r},\mathcal{t}}$ \;
}
\end{algorithm}

\subsubsection{Conversation Module.}
Conversation module encodes current state and make decisions, including asking attributes or making recommendation, according to the current state. The policy network, which is used to make decisions, is a multi-layer perceptron and optimized via the policy gradient method of reinforcement learning \cite{sutton2018reinforcement}. We will first introduce the design of the state vector in our KG-CRS, and then explain the details of the policy network.
It is to be noted, as we have stated, graph $\mathcal{G}$ is dynamic during the conversation process, which might directly affect the result of Equation \ref{eq:final embedding}, and then further influence the recommendation results and the state vector in conversation module. In particular, in triples $(\mathcal{h},\mathcal{r},\mathcal{t})$, negative entities in $P_{neg}$ and $V_{neg}$ are removed from $\mathcal{G}$, so that the information of these negative entities will not influence the recommendation and conversational results. For example, considering two candidate items $v_1$ and $v_2$ in the current session, and item $v_1$ is much more similar to item $v_3$ than $v_2$. On the other hand, item $v_3$ was chosen by user $u$ in the history, but rejected by $u$ in the current session. In this case, if item $v_3$ is still involved in the graph $\mathcal{G}$, item $v_1$ may get a higher score than $v_2$ because it is more similar to item $v_3$ in the history. Therefore, in order to address this issue, we update graph $\mathcal{G}$ each turn by removing the negative entities (e.g., item $v_3$) and resume it for every new session. In this example, we only remove item $v_3$ but its related attributes will still exist in the graph. 
Consequently, for every related attribute $p$, it only loses one relation with $v_3$ in embedding learning, but can still propagate its information by other relations.

\medskip\textbf{State Vector.} As defined in Equation \ref{eq:state_vector}, state vector consists of four components, entropy-based probability $\textbf{s}_{ent}$, user-based probability $\textbf{s}_{user}$, conversation-based probability $\textbf{s}_{conv}$ and dialogue feature $\textbf{s}_{dial}$. The first three parts calculate the probability of which attribute to ask from three different perspectives. The fourth part records the situation of current conversation, which is expected to guarantee the learning on whether to recommend or ask.
\begin{equation} \label{eq:state_vector}
    \textbf{s} = [\textbf{s}_{ent}, \textbf{s}_{user}, \textbf{s}_{conv}, \textbf{s}_{dial}]
\end{equation}
where $\textbf{s}_{ent}$ is a vector composed of the entropy of all attributes, whose size equals to the size of attributes. The entropy information of each attribute is calculated as Equation \ref{eq:entropy}, which will be changed as the candidate items change. The agent can narrow the candidate set of items quickly through asking attributes with large entropy. This vector is one component of the state vector in \cite{lei2020estimation}, and we keep it as it has a good ability for the selection of attributes to ask.
\begin{equation} \label{eq:entropy}
    H(x)=-\sum_xp(x)\log p(x)
\end{equation}
where $p(x)$ represents the probability of attribute $x$ in the set of candidate items.

$\textbf{s}_{user}$ is a vector composed of the scores of all attributes, and the size of the vector is equivalent to the number of attributes.
The score of an attribute $p$, $y_p^u$, is calculated according to the embedding of user $u$ and attribute $p$, as shown in Equation \ref{eq:user-based-prob}. The corresponding embedding, $\textbf{u}_{2}$ and $\textbf{p}_{2}$, is obtained via graph embedding learning. That is, $\textbf{u}_{2}= M_{\mathcal{2u}}\textbf{u}$, $\textbf{p}_{2}= M_{\mathcal{2p}}\textbf{p}$, where \textbf{u}, \textbf{p} are the embedding of user $u$ and attribute $p$ measured using Equation \ref{eq:final embedding}, and $M_{\mathcal{2u}}$, $\ M_{\mathcal{2p}}$ are the projection matrices in terms of the user-attribute relation in Equation \ref{eq:projection_matrix}.
The motivation of component $\textbf{s}_{user}$ is that user $u$ may keep the same preference as demonstrated in her historical interactions. As the graph embedding is pre-trained before the conversation module (see Algorithm \ref{alg:embedding learning}), attributes that user $u$ often care will obtain a higher score, which can guide the policy network to select more appropriate attributes.
\begin{equation} \label{eq:user-based-prob}
    y_p^u = \textbf{u}_{2}^T\textbf{p}_{2}
\end{equation}

$\textbf{s}_{conv}$ is also a vector composed of the scores of all attributes while the score of each attribute $p$, $y_p^c$, is calculated according to the embedding of all known positive attributes $P_{u}$ and candidate attribute $p$, as shown in Equation \ref{eq:conv-based-prob}. Specifically, ${\textbf{p}_{u}}_{1}= M_{\mathcal{1p}}\textbf{p}_{u}$, $\textbf{p}_{1}= M_{\mathcal{1p}}\textbf{p}$, where $\textbf{p}_{u}$, \textbf{p} are the embedding of attribute $p_u$ and attribute $p$ calculated using Equation \ref{eq:final embedding}, and $\ M_{\mathcal{1p}}$ are the projection matrices in terms of item-attribution relation in Equation \ref{eq:projection_matrix}. 
$P_{u}$ represents the set of attributes that has been confirmed by the user in the previous rounds of the conversation session. The objective of this component is to guide the agent on selecting an attribute to ask based on the immediate user preference in the session. That is, the more relevant $p_u$ and $p$ are, the higher the score is. Its size is still the same as the size of attributes.
\begin{equation} \label{eq:conv-based-prob}
    y_p^c = \sum_{p_{u} \in P_{u}}{\textbf{p}_{u}}_{1}^T\textbf{p}_{1}
\end{equation}

 $\textbf{s}_{dial}$ is a vector that encodes the situation of current conversation, which mainly considers the length of candidate items and the dialogue history.
 In this work, we follow its design as \cite{lei2020estimation}, and leave its variant as our future work. Specifically, the length of candidate items is encoded as a one-hot vector with the motivation that it is easy for the agent to recommend successfully when the length of candidate items is short, unsuccessfully otherwise. The dialogue history is represented as a fixed-length (i.e., the maximum number of turns $T$) vector and each number in this vector represents the situation of the corresponding dialogue turn, where $1$ means asking a positive attribute, $0$ means asking an irrelevant attribute, and $-1$ means making an unsuccessful recommendation. The number at the position exceeding the current dialogue turn $t$ is also represented as 0.
 
\medskip\textbf{Reward.} After the agent takes an action according to current state, user will respond and correspondingly get a numerical reward.
As shown in Table \ref{tb:rewards}, user $u$'s responses can be concluded into five categories, and we specifically adopt two different designs of the reward function (i.e., coarse-grained reward (CG reward) and fine-grained reward (FG reward)) to address these responses.

\begin{table}[htb]
\centering
\small
  \caption{Reward function design for different user response.}
  \label{tb:rewards}
  \begin{tabular}{l|ll}
    \hline
    user response at turn $t$ & CG reward $r_t$ & FG reward $r_t$\\
    \hline
    \# accept recommendation & $r_{item}$ & $r_{item}$\\
    \# relevant questions about attributes & $r_{attr}$ + $r_{turn}$ & $r_{attr}$ + $r_{turn}$ + $r_{imp}$\\
    \# irrelevant questions about attributes & $r_{turn}$ & $r_{turn}$\\
    \# reject recommendation & $r_{turn}$ & $r_{turn}$ \\
    \# user quit & $r_{quit}$ & $r_{quit}$ \\
  \hline
\end{tabular}
\end{table}

The design of the coarse-grained reward is similar to the design in \cite{lei2020estimation}, which consists of four parts with the objective to find a policy network which can lead to successful recommendation but reduce irrelevant questions. Specifically, $r_{attr}$ is a small positive reward when the question about attribute is positively answered. $r_{item}$ represents a positive reward when the recommendation is successful. $r_{turn}$ denotes a small negative reward for each turn (round) to penalize when the number of rounds in this session gets longer. $r_{quit}$ is a negative reward when the user quits the conversation.

In our paper, we also design the fine-grained reward framework which aims to further minimize the number of rounds/turns in a session by reducing relevant but unnecessary questions with $r_{imp}$. The intuition of introducing $r_{imp}$ is that, if user $u$ consistently interacts with items having attribute $p$, the system is likely to capture this behavior. In this case, it might be unnecessary for the CRS agent to ask $u$ about attribute $p$.
And, the soundness of this intuition is also dependent on whether the recommendation module has precisely learned preference of user $u$. 
Therefore, under these considerations, $r_{imp}$ is designed as shown in Equation \ref{eq:action effect}. Specifically, $loc$ refers to the ideal item position in a turn, where the ideal item means a ground-truth item user $u$ will interact with, and its position means the ranking order given by the recommendation module. $loc_t$ means the ideal item position before action $a_t$ (the action conducted at turn $t$), while $loc_{t+1}$ refers to the position after $a_t$. In this case, $\frac{loc_{t} - loc_{t+1}} { loc_{t}}$ measures the impact of action $a_t$. Namely, a more necessary question can lead to greater position change of the idea item. $\beta$ is a hyper-parameter that balances the impact of $r_{imp}$ and other rewards, while the fine-grained reward framework reduces to the coarse-grained reward with $\beta=0$.
\begin{equation} \label{eq:action effect}
    r_{imp} = \beta * \frac{loc_{t} - loc_{t+1}} { loc_{t}}
\end{equation}

\medskip\textbf{Policy Network.} The policy network is a multi-layer perceptron, which takes the state vector \textbf{s} as the input and outputs a probability distribution of all actions. The dimension of the probability distribution is $N_q + 1$. The reward $r_t$, as shown in Table \ref{tb:rewards}, is the summation of all related rewards at turn $t$, and the policy network is optimized according to Equation \ref{eq:reinforcement_learning}.
\begin{equation} \label{eq:reinforcement_learning}
    \theta \gets \theta + \eta \nabla_{\theta} \log \pi_{\theta}(a_t|s_t)\sum_{t'=t, ..., T'} \gamma^{(t'-t)}r_{t'} 
\end{equation}
where $\theta$ represents the policy network's parameter, $\eta$ is the learning rate, $\pi_{\theta}(a_t|s_t)$ denotes the probability of action $a_t$ given state $s_t$, and $\gamma$ is a discount factor. $T'$ is the total number of turns in this session.
The training process of the entire conversation module is detailed in Algorithm \ref{alg:conversation module}.

\begin{algorithm}
\small
\caption{Training of Conversation Module}
\label{alg:conversation module}
\KwData{all attributes $P$, all items $V$, all actions $A$, the length of recommend items list $K$, the set of pretrained embedding $E_{\mathcal{h},\mathcal{r},\mathcal{t}}$, the set of pretrained parameters $W$ and $b$, propagation step $\mathcal{K}$, the maximum number of rounds $T$, the policy network $\theta$}
\For{\emph{user} $u$ \emph{and his preference} $P_{ses}$ \emph{in this session \textbf{in}} training data}{
    $\mathcal{G}^{'} \leftarrow \mathcal{G}$; $E_{\mathcal{h},\mathcal{r},\mathcal{t}}^{'} \leftarrow E_{\mathcal{h},\mathcal{r},\mathcal{t}}$; $W' \leftarrow W$, $b' \leftarrow b$ \;
    user $u$ specifies an attribute $p$\;
    $t \leftarrow 1$; $P_{u} \leftarrow \{p\}$; $P_{neg} \leftarrow \{\}$; $V_{cand} \leftarrow V_{p}$; $V_{neg} \leftarrow \{\}$; $A_{cand} \leftarrow A$; $A_{know} \leftarrow \{a_{p}\}$; $r_{list} \leftarrow []$\;
    \While{$t\leq T$}{
        calculate the score of items in $V_{cand}$ with graph $\mathcal{G}^{'}$ and parameters $E_{\mathcal{h},\mathcal{r},\mathcal{t}}^{'}$, $W'$, $b'$\;
        calculate $\textbf{s}_{t}$ with graph $\mathcal{G}^{'}$ and parameters $E_{\mathcal{h},\mathcal{r},\mathcal{t}}^{'}$, $W'$, $b'$ \;
        sample an action $a_t$ based on $\pi_{\theta}(\textbf{s}_t)$, $a_t \in A_{cand}$ \;
        \eIf(\tcp*[f]{agent decides to ask attributes}){$a_t \in A_{ask}$}{
            $A_{know} \leftarrow A_{know} + \{a_t\}$; $A_{cand} \leftarrow A_{cand} - \{a_t\}$\;
            \eIf(\tcp*[f]{user directly informs the result of intersection}){$P_{a_t} \cap P_{ses} \neq \varnothing$}{
                $P_u \leftarrow P_u + P_{a_t} \cap P_{ses}$; $P_{neg} \leftarrow P_{neg} + P_{a_t} - P_{a_t} \cap P_{ses}$; \tcp*{user care}
                $V_{cand} \leftarrow V_{cand} \cap V_{P_{a_t} \cap P_{ses}}$ \;
                $r_t \leftarrow r_{attr} + r_{turn} + r_{imp}$, append $r_t$ to $r_{list}$\;
            }
            {
                $P_{neg} \leftarrow P_{neg} + P_{a_t}$ \tcp*{user does not care}
                $r_t \leftarrow r_{turn}$, append $r_t$ to $r_{list}$\;
            }
        }
        (\tcp*[f]{agent decides to make recommendation})
        {
            select top-$K$ items $V_{rec}$, $V_{rec} \subseteq V_{cand}$\;
            \eIf{\emph{user} $u$ \emph{accepts recommendation}}{
                \textbf{break} \tcp*{user accepts recommendation}
            }
            {
                $V_{cand} \leftarrow V_{cand}-V{rec}$; $V_{neg} \leftarrow V_{neg}+V_{rec}$ \tcp*{user rejects recommendation}
                $r_t \leftarrow r_{turn}$, append $r_t$ to $r_{list}$\;
                \textbf{update} parameters $E_{\mathcal{h},\mathcal{r},\mathcal{t}}^{'}$, $W'$, $b'$ with positive items of $u$ and $V_{neg}$ as line 9-17 in Algorithm \ref{alg:embedding learning}
            }
        }
        \textbf{update} graph $\mathcal{G}^{'}$, remove triple $(\mathcal{h},\mathcal{r},\mathcal{t})$ that contain entities in $P_{neg}$ and $V_{neg}$\ from graph $\mathcal{G}^{'}$;
    }
    \eIf{$t\leq T$}{
        $r_t \leftarrow r_{item}$, append $r_t$ to $r_{list}$\;
    }{
        $r_t \leftarrow r_{quit}$, append $r_t$ to $r_{list}$\;
    }
    \textbf{update} policy network $\theta$ with $L_{conv}$, $L_{conv}= - \sum_{t = 1, ..., len(r_{list})} \log \pi_{\theta}(a_t|s_t)\sum_{t' = t, ..., len(r_{list})} \gamma^{(t'-t)}r_{t'}$\;
}
\end{algorithm}

\subsection{Time Complexity Analysis}
The main operations in our method are listed as follows:
\begin{itemize}
\item \textbf{Operation 1: node feature learning.} It locates in lines 5-7 in Algorithm \ref{alg:embedding learning}
\item \textbf{Operation 2: information propagation and score calculation.} It lies in lines 9-16 in Algorithm \ref{alg:embedding learning} and line 6 in Algorithm \ref{alg:conversation module}.
\item \textbf{Operation 3: updating weight $\alpha_{\mathcal{h},\mathcal{r},\mathcal{t}}$.} It refers to lines 2 and 19 in Algorithm \ref{alg:embedding learning}.
\item \textbf{Operation 4: encoding state vector.} This operation locates in line 7 in Algorithm \ref{alg:conversation module}.
\item \textbf{Operation 5: taking action.} It refers to line 8 in Algorithm \ref{alg:conversation module}.
\item \textbf{Operation 6: updating graph structure (i.e., removing negative items and negative attributes).} It lies in line 29 in Algorithm \ref{alg:conversation module}.
\end{itemize}

The time complexity of these operations are summarized in Table \ref{tb:complexity}, where $N=|U|+|V|+|P|$, $m$ is the embedding size, $n$ is the batch size, $n_{hid}$ is the hidden size of the policy network and other notations are shown in Table \ref{tb:notations}. The experiments are conducted with NVIDIA Tesla K80 and 2 CPU core (2.20GHZ).

\begin{table*}[htb]
\small
\centering
  \caption{Time Complexity Analysis.}
  \label{tb:complexity}
  \begin{tabular}{l|l|lll}
    \hline
    & & \multicolumn{3}{c}{Running Time(s)}\\
    \cline{3-5}
    Operation & Time Complexity & Yelp & LastFM & Amazon\\
    \hline
    \# 1 & $\mathcal{O}(nm)$ & 0.1707 & 0.0061 & 0.0057\\
    \# 2 & $\mathcal{O}((\mathcal{K}Nm+n)(1+m))$ & 0.3608 & 0.0321 & 0.0153\\
    \# 3 & $\mathcal{O}(Nm(1+N))$ & 170.3350 & 15.0842 & 7.9716\\
    \# 4 & $\mathcal{O}((\mathcal{K}Nm+|P|)(1+m)+|P|(1+|V_{cand}|))$ & 0.4275 & 0.0525 & 0.0184\\
    \# 5 & $\mathcal{O}(n_{hid}(len(\textbf{s})+|A|))$ & 0.0009 & 0.0015 & 0.0009\\
    \# 6 & $\mathcal{O}(N^2(|P_{neg}|+|V_{neg}|))$ & 0.0828 & 0.0170 & 0.0056\\
  \hline
\end{tabular}
\end{table*}

\section{Experiments}
\label{sec:experiments}
In this section, we conduct experiments on three real-world datasets to validate the effectiveness of our proposed KG-CRS, with the goal of answering three research questions:
\begin{itemize}
\item \textbf{RQ1:} How does our KG-CRS perform compared with other state-of-the-art approaches in terms of both recommendation task and conversation task?

\item \textbf{RQ2:} How do different components of KG-CRS contribute to the performance?

\item \textbf{RQ3:} How do different hyper-parameters affect the performance of KG-CRS?

\end{itemize}
\subsection{Experiment Settings}
\subsubsection{Datasets.}
We conduct experiments on three datasets: Yelp\footnote{\url{www.yelp.com/dataset/}.}, LastFM\footnote{\url{grouplens.org/datasets/hetrec-2011/}.} and Amazon\footnote{\url{jmcauley.ucsd.edu/data/amazon/}.}. 
All these datasets provide rich information about users and items, including user's interaction history with items, item's price, and item's categories. In this work, we utilize the interaction history between users and items, and extract items' attributes from the rich information.
For Yelp and LastFM, we follow the similar data processing as \cite{lei2020estimation}. Specifically, users that have less than $10$ interactions are removed from datasets to reduce the data sparsity. Besides, the tag data regarding items on LastFM is directly used as items' attributes, while items' attributes on Yelp are a combination of various features towards items, including items' city, price and category information. 
For Amazon, we mainly explore one of the sub-datasets, i.e., ``Pet Supplies'', which has the same sets of users and items as \cite{zou2020towards}, and users and items that have less than $5$ interactions are removed. Following \cite{zou2020towards}, items' attributes are extracted from review text using Gensim \cite{rehurek2010software}. 
%All these attributes in the three datasets are filtered by frequency and synonyms.
Each dataset is divided into training/valid/test sets in the ratio of 7:2:1. The statistics of the three datasets are summarized in Table \ref{tb:datasets}.

\begin{table}[t]
\centering
%\small
  \caption{Statistics of the two datasets.}
%  \vspace{-0.1in}
  \label{tb:datasets}
  \begin{tabular}{l|ccc}
    \hline
    Dataset& \textbf{Yelp}& \textbf{LastFM}& \textbf{Amazon}\\
    \hline
    \# users & 27,675 & 1,801 & 2,248\\
    \# items & 70,311 & 7,432 & 2,475\\
    \# interactions & 1,368,606 & 76,693 & 13,939\\
   \# attributes& 590 & 33 & 115\\
  \hline
\end{tabular}
\end{table}

\subsubsection{User Simulator and Conversation Setting.} As our CRS is dynamic and the consideration of real users is costly, we follow \cite{lei2020estimation} to create a user simulator to replace real users for the model training and testing. The user simulator randomly samples an attribute of the ideal item as the beginning of the conversation and gives a response according to the question or the recommendation in the following turns. We use the multi-round recommendation setting, which means that the user will not quit when the recommendation fails until the predefined maximum number of rounds $T$ reaches. Same as \cite{lei2020estimation, lei2020interactive}, the agent will recommend $K$ items (i.e., $K=10$) each time. $T$ is set as $15$ if not specifically identified in the following experiments. 

As illustrated in \cite{lei2020estimation}, there are generally two kinds of questions in CRS: binary question and enumerated question, which are also referred to as yes/no question and open question in some other literature \cite{xu2020user} respectively. The former one may ask questions like `Are you in Washington?' while the latter one may ask like `Which city are you in?'. Platforms can determine to use which kind of question setting according to their engineering and business needs. For fair comparisons, we follow \cite{lei2020estimation} to test our model with both two settings. Specifically, the experiments on Yelp is conducted with enumerated question, while on LastFM and Amazon is binary question. Moreover, $590$ attributes on Yelp are categorized into $29$ facets given the enumerated setting, which means $N_q$ on Yelp is equivalent to $29$, namely the action space equals to $30$. Given the binary setting on LastFM and Amazon, one question is related to one attribute and $N_q$ on LastFM and Amazon thus equals $33$ and $115$ respectively, and thus the action space on is equal to $34$ and $116$ respectively.

\subsubsection{Baselines.} 
Although there are numerous works \cite{christakopoulou2016towards, lei2020estimation, lei2020interactive, sun2018conversational, zou2020towards, zhang2018towards, li2018towards, zhou2020improving, 48414, xu2020user} on CRS, as discussed in Section \ref{sec:relatedwork}, many of them \cite{zou2020towards, zhang2018towards, li2018towards, zhou2020improving, 48414, xu2020user} have different settings or focuses from our study. For example, \cite{48414} constructed a system which interacts with users like a chat-bot rather than a task-oriented system. \cite{li2018towards, zhou2020improving} focused on the NLP and semantic information. Thus, we compare our performance with the following most related baselines since their settings and research focus are much closer to ours:
\begin{itemize}
\item \textbf{Abs Greedy} \cite{christakopoulou2016towards}: The main idea of the dialogue strategy for this method is to keep making recommendations without asking any questions. To better compare this rule-based dialogue strategy with the trained policy network in KG-CRS, the recommendation module in this method is the same as that in KG-CRS.
\item \textbf{Max Entropy} \cite{sun2018conversational}: This method is also designed for the dialogue strategy and chooses to ask the attribute that has the maximum entropy among the candidates in each turn. The recommendation is randomly triggered by the length of the candidate item list. While the length gets shorter, it becomes more likely to do the recommendation. 
Similar to Abs Greedy, the recommendation module in this method is also the same as that in KG-CRS.
\item \textbf{CRM} \cite{sun2018conversational}: The dialogue strategy in this method is decided by a neural network using a reinforcement learning framework, and the recommendation module is fulfilled by Factorization Machine(FM). It is originally designed for the single-round recommendation scenario. We adapt it to the multi-round recommendation scenario by following the process of \cite{lei2020estimation}.
\item \textbf{EAR} \cite{lei2020estimation}: Similar to CRM, the conversation module and the recommendation module of this method also rely on reinforcement learning and FM respectively. Different from CRM, it is applied in the multi-round setting and proposes a three-stage solution called Estimation-Action-Reflection framework which updates the recommendation module according to the conversation module.
\item \textbf{SCPR} \cite{lei2020interactive}: This is the state-of-the-art method on CRS under the multi-round setting. It is built on EAR and further considers the graph-based conversational path reasoning to help narrow the candidate set of attributes and items.
\end{itemize}

\subsubsection{Parameters Setup.} 

\begin{figure}[t]
\centering
\includegraphics[scale=0.23]{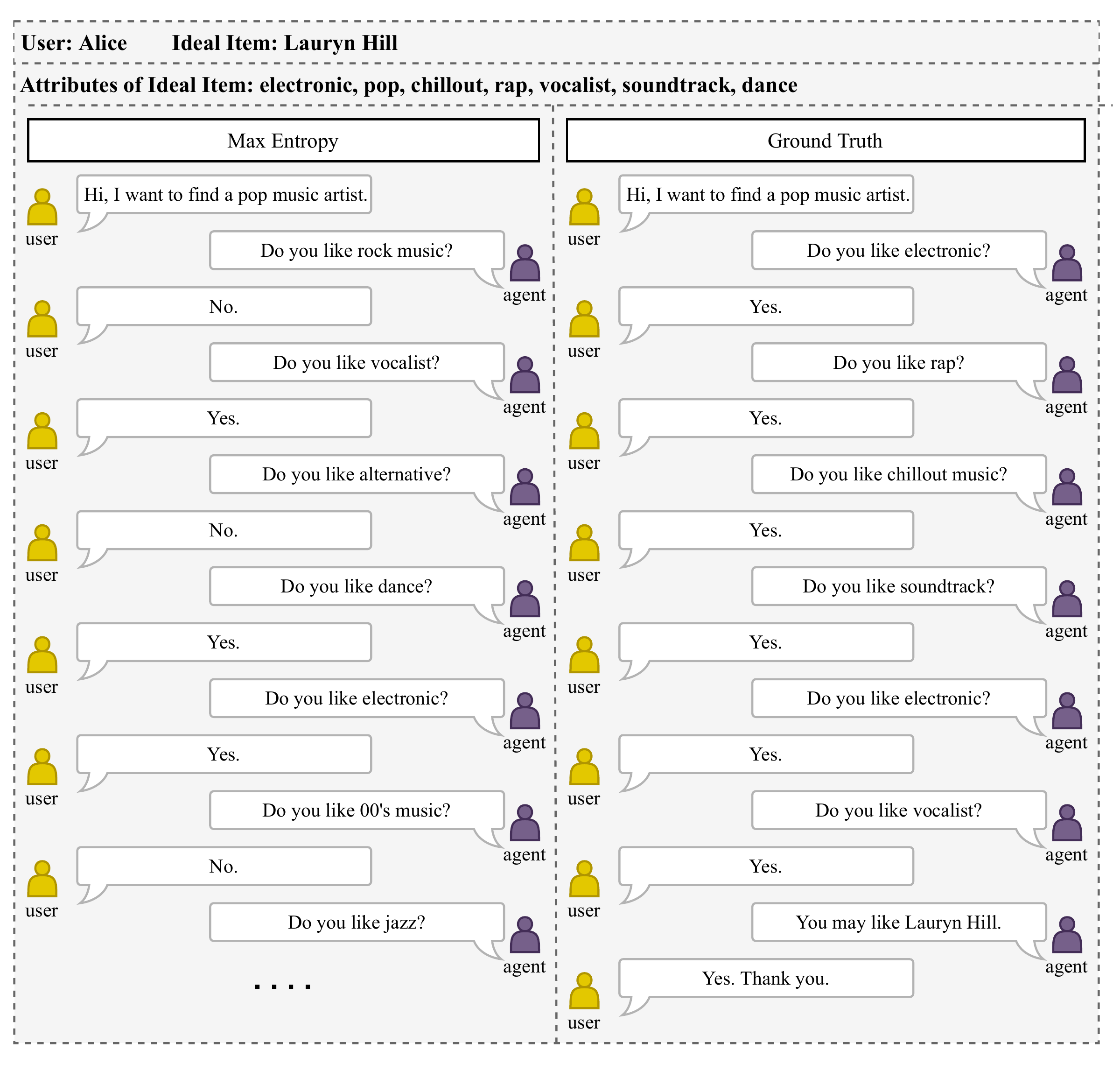}
\caption{Examples depicting the two pre-trained strategies.}
\label{fig:pretrain}
\end{figure}

Following \cite{sun2018conversational, lei2020estimation}, the parameters of our framework is initialized with offline data firstly and then optimized online through the user simulator. The parameters of graph embedding can be pre-trained using the graph. It is worth mentioning that there is no available conversation data on Yelp, LastFM and Amazon for initializing the parameters of the policy network. The studies of \cite{zhang2020conversational, lei2020estimation} choose to let the policy network mimic the strategy of Max Entropy, whereas in our KG-CRS, besides the Max-Entropy strategy, we also propose to let the policy network mimic the ground truth of this problem. Specifically, in policy network pretrained phase, as shown in Figure \ref{fig:pretrain}, assuming that user `Alice' wants to find item `Lauryn Hill' through the CRS, it is better if the agent can ask attributes relating to `Lauryn Hill' precisely and then make the recommendation. As the unstable performance of reinforcement learning, we consider this strategy can better initialize the parameters for the policy network in KG-CRS.

In KG-CRS, in graph embedding and recommendation module, the propagation step and embedding size of each step are set as $4$ and $64$ on Yelp, LastFM and Amazon, while the corresponding learning rate is $0.001$ with the SGD optimizer. Following the parameter setting in \cite{han2018openke}, the margin parameter $\lambda$ in Equation \ref{eq:loss_graph} is set as $4$.
In the conversation module, on all datasets, if not specified, we mainly adopt the coarse-grained reward framework, and the rewards of $r_{item}$, $r_{attr}$, $r_{turn}$, and $r_{quit}$ are set as $1$, $0.1$, $-0.01$, and $-0.3$, respectively. $\beta$ is set as $0.1$ for the fine-grained reward. The discount factor $\gamma$ in Equation \ref{eq:reinforcement_learning} is set as $0.7$. The learning rate and optimizer on Yelp are $0.001$ and SGD respectively, while these two hyper-parameters on LastFM and Amazon are $0.00001$ and Adam respectively. 

For CRM, EAR, and SCPR, we explicitly adopt their released pre-trained parameters and hyper-parameters to conduct experiments.

\subsubsection{Evaluation Metrics.} 
For evaluating the performance on recommendation task, we adopt the success rate (SR@T) \cite{sun2018conversational,lei2020estimation} to measure the ratio of conversations which have successfully recommended the ground-truth items by turn $T$. We also specifically compare the performance of different approaches' recommendation modules on offline dataset using Precision@K, Recall@K, NDCG@K, and AUC (area under ROC curve) metric.
For evaluating the performance on conversation task, similar to \cite{lei2020estimation}, we use the average turns (AT) needed to end the conversation-recommendation process.
Besides, we also propose a new metric, the average positive action (APA) (either asking a favorable attribute or providing successful recommendation), to measure users' satisfaction towards the conversation process.
For SR, APA, Precision@K, Recall@K, NDCG@K, and AUC, a larger number indicates better performance while smaller AT means more efficient CRS.

\subsection{Experimental Results}
Here, we present our experimental results towards the aforementioned research questions.
\subsubsection{Comparative Results (RQ1).}

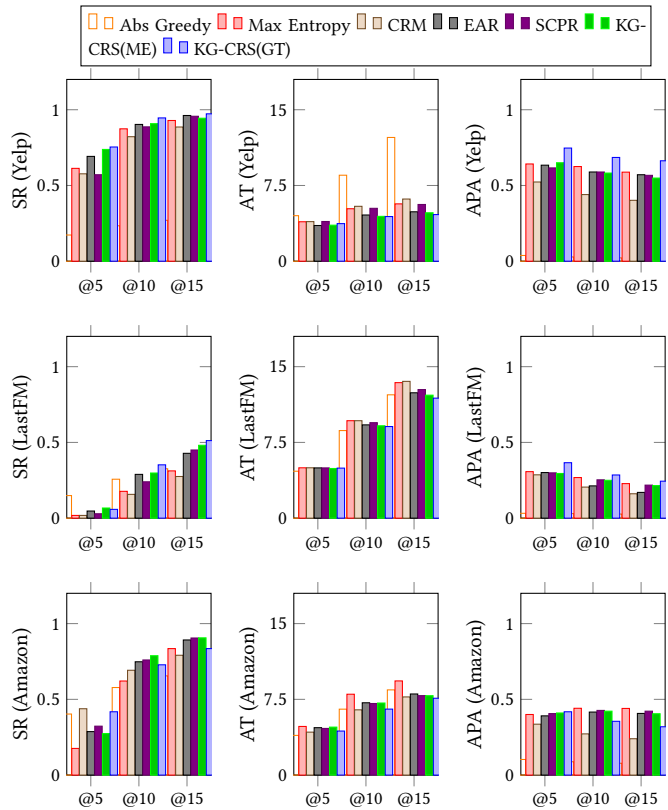
\begin{figure}[htb]
    \centering
	\footnotesize
	\begin{tikzpicture}
	\begin{groupplot}[group style={
		group name=myplot,
		group size= 3 by 3,  horizontal sep=1.1cm}, 
		height=4cm, width=3.5cm,
	ylabel style={yshift=-0.1cm},
	legend style = {font=\small, column sep=-1cm},
	every tick label/.append style={font=\footnotesize}
	]
	
	% SR, Yelp
	\hspace{-0.2in}
	\nextgroupplot[ybar=0.10,
	bar width=0.4em,
	ylabel={SR (Yelp)},
	scaled ticks=false,
	yticklabel style={/pgf/number format/.cd,fixed,precision=3},
	ymin=0, ymax=1.2,
	enlarge x limits=0.25,
	symbolic x coords={@5, @10, @15},
	ylabel style = {font=\small},
	xtick=data,
	ytick={0,0.5,1.0},
	]
	\addplot[color=orange] coordinates {
		(@5,0.173) (@10, 0.234) (@15, 0.270)};\label{plots:tplot1}
	\addplot coordinates {
		(@5,0.613) (@10, 0.874) (@15, 0.929)};\label{plots:tplot2}
	\addplot coordinates {
		(@5,0.577) (@10, 0.822)  (@15, 0.886)};\label{plots:tplot3}
	\addplot coordinates {
		(@5,0.692) (@10, 0.903) (@15, 0.962)};\label{plots:tplot4}
	\addplot coordinates {
		(@5,0.571) (@10, 0.887)  (@15, 0.957)};\label{plots:tplot5}
	\addplot coordinates {
		(@5,0.738) (@10, 0.908)  (@15, 0.944)};\label{plots:tplot6}
	\addplot coordinates {
		(@5,0.754) (@10, 0.946)  (@15, 0.973)};\label{plots:tplot7}

	% AT, Yelp
	\nextgroupplot[ybar=0.1,
	bar width=0.4em,
	ylabel={AT (Yelp)},
	scaled ticks=false,
	yticklabel style={/pgf/number format/.cd,fixed,precision=3},
	ymin=0, ymax=18,
	enlarge x limits=0.25,
	symbolic x coords={@5,@10, @15},
	xtick=data,
	ylabel style = {font=\small},
	ytick={0, 7.5, 15},]
	\addplot[color=orange] coordinates {
		(@5,4.509) (@10, 8.516) (@15, 12.255)};
	\addplot coordinates {
		(@5,3.908) (@10, 5.197) (@15, 5.679)};
	\addplot coordinates {
		(@5,3.921) (@10, 5.437) (@15, 6.157)};
	\addplot coordinates {
		(@5,3.537) (@10, 4.577) (@15, 4.894)};
	\addplot coordinates {
		(@5,3.934) (@10, 5.239) (@15, 5.619)};
    \addplot coordinates {
		(@5,3.570) (@10, 4.443) (@15, 4.798)};
	\addplot coordinates {
		(@5,3.723) (@10, 4.423) (@15, 4.618)};
		
	% APA, Yelp
	\nextgroupplot[ybar=0.1,
	bar width=0.4em,
	ylabel={APA (Yelp)},
	scaled ticks=false,
	yticklabel style={/pgf/number format/.cd,fixed,precision=3},
	ymin=0, ymax=1.2,
	enlarge x limits=0.25,
	symbolic x coords={@5, @10, @15},
	xtick=data,
	ylabel style = {font=\small},
	ytick={0, 0.5, 1.0},]
    \addplot[color=orange] coordinates {
    (@5, 0.038) (@10, 0.028) (@15, 0.022)};
    \addplot coordinates {
    (@5, 0.642) (@10, 0.625) (@15, 0.588)};
    \addplot coordinates {
    (@5, 0.523) (@10, 0.440) (@15, 0.402)};
    \addplot coordinates {
    (@5, 0.634) (@10, 0.589) (@15, 0.571)};
    \addplot coordinates {
    (@5, 0.616) (@10, 0.589) (@15, 0.567)};
    \addplot coordinates {
    (@5, 0.650) (@10, 0.582) (@15, 0.549)};
    \addplot coordinates {
    (@5, 0.747) (@10, 0.685) (@15, 0.663)};
    
	% SR, LastFM
	\nextgroupplot[ybar=0.10,
	bar width=0.4em,
	ylabel={SR (LastFM)},
	scaled ticks=false,
	yticklabel style={/pgf/number format/.cd,fixed,precision=3},
	ymin=0, ymax=1.2,
	enlarge x limits=0.25,
	symbolic x coords={@5, @10, @15},
	ylabel style = {font=\small},
	xtick=data,
	ytick={0,0.5,1.0},
	]
	\addplot[color=orange] coordinates {
		(@5,0.149) (@10, 0.257) (@15, 0.326)};
	\addplot coordinates {
		(@5,0.018) (@10, 0.177) (@15, 0.312)};
	\addplot coordinates {
		(@5,0.018) (@10, 0.157)  (@15, 0.275)};
	\addplot coordinates {
		(@5,0.047) (@10, 0.289) (@15, 0.428)};
	\addplot coordinates {
		(@5,0.029) (@10, 0.240)  (@15, 0.450)};
	\addplot coordinates {
		(@5,0.067) (@10, 0.298)  (@15, 0.481)};
	\addplot coordinates {
		(@5,0.058) (@10, 0.352)  (@15, 0.512)};

	% AT, LastFM
	\nextgroupplot[ybar=0.1,
	bar width=0.4em,
	ylabel={AT (LastFM)},
	scaled ticks=false,
	yticklabel style={/pgf/number format/.cd,fixed,precision=3},
	ymin=0, ymax=18,
	enlarge x limits=0.25,
	symbolic x coords={@5,@10, @15},
	xtick=data,
	ylabel style = {font=\small},
	ytick={0, 7.5, 15},]
	\addplot[color=orange] coordinates {
		(@5,4.647) (@10, 8.672) (@15, 12.213)};
	\addplot coordinates {
		(@5,4.985) (@10, 9.650) (@15, 13.419)};
	\addplot coordinates {
		(@5,4.986) (@10, 9.649) (@15, 13.545)};
	\addplot coordinates {
		(@5,4.975) (@10, 9.237) (@15, 12.414)};
	\addplot coordinates {
		(@5,4.976) (@10, 9.454) (@15, 12.732)};
    \addplot coordinates {
		(@5,4.918) (@10, 9.147) (@15, 12.178)};
	\addplot coordinates {
		(@5,4.954) (@10, 9.072) (@15, 11.883)};
		
	% APA, LastFM
	\nextgroupplot[ybar=0.1,
	bar width=0.4em,
	ylabel={APA (LastFM)},
	scaled ticks=false,
	yticklabel style={/pgf/number format/.cd,fixed,precision=3},
	ymin=0, ymax=1.2,
	enlarge x limits=0.25,
	symbolic x coords={@5, @10, @15},
	xtick=data,
	ylabel style = {font=\small},
	ytick={0, 0.5, 1.0},]
    \addplot[color=orange] coordinates {
    (@5, 0.032) (@10, 0.030) (@15, 0.027)};
    \addplot coordinates {
    (@5, 0.307) (@10, 0.268) (@15, 0.228)};
    \addplot coordinates {
    (@5, 0.286) (@10, 0.205) (@15, 0.161)};
    \addplot coordinates {
    (@5, 0.301) (@10, 0.213) (@15, 0.170)};
    \addplot coordinates {
    (@5, 0.299) (@10, 0.253) (@15, 0.218)};
    \addplot coordinates {
    (@5, 0.295) (@10, 0.250) (@15, 0.215)};
    \addplot coordinates {
    (@5, 0.366) (@10, 0.285) (@15, 0.244)};
    
	% SR, Amazon
	\nextgroupplot[ybar=0.10,
	bar width=0.4em,
	ylabel={SR (Amazon)},
	scaled ticks=false,
	yticklabel style={/pgf/number format/.cd,fixed,precision=3},
	ymin=0, ymax=1.2,
	enlarge x limits=0.25,
	symbolic x coords={@5, @10, @15},
	ylabel style = {font=\small},
	xtick=data,
	ytick={0,0.5,1.0},
	]
	\addplot[color=orange] coordinates {
		(@5,0.403) (@10, 0.578) (@15, 0.655)};
	\addplot coordinates {
		(@5,0.176) (@10, 0.621) (@15, 0.835)};
	\addplot coordinates {
		(@5,0.438) (@10, 0.692)  (@15, 0.791)};
	\addplot coordinates {
		(@5,0.287) (@10, 0.748) (@15, 0.892)};
	\addplot coordinates {
		(@5,0.323) (@10, 0.760)  (@15, 0.905)};
	\addplot coordinates {
		(@5,0.274) (@10, 0.788)  (@15, 0.906)};
	\addplot coordinates {
		(@5,0.418) (@10, 0.728)  (@15, 0.835)};

	% AT, Amazon
	\nextgroupplot[ybar=0.1,
	bar width=0.4em,
	ylabel={AT (Amazon)},
	scaled ticks=false,
	yticklabel style={/pgf/number format/.cd,fixed,precision=3},
	ymin=0, ymax=18,
	enlarge x limits=0.25,
	symbolic x coords={@5,@10, @15},
	xtick=data,
	ylabel style = {font=\small},
	ytick={0, 7.5, 15},]
	\addplot[color=orange] coordinates {
		(@5,3.930) (@10, 6.527) (@15, 8.437)};
	\addplot coordinates {
		(@5,4.816) (@10, 8.007) (@15, 9.324)};
	\addplot coordinates {
		(@5,4.252) (@10, 6.460) (@15, 7.739)};
	\addplot coordinates {
		(@5,4.694) (@10, 7.160) (@15, 8.031)};
	\addplot coordinates {
		(@5,4.620) (@10, 7.088) (@15, 7.867)};
    \addplot coordinates {
		(@5,4.750) (@10, 7.136) (@15, 7.861)};
	\addplot coordinates {
		(@5,4.359) (@10, 6.532) (@15, 7.609)};
		
	% APA, Amazon
	\nextgroupplot[ybar=0.1,
	bar width=0.4em,
	ylabel={APA (Amazon)},
	scaled ticks=false,
	yticklabel style={/pgf/number format/.cd,fixed,precision=3},
	ymin=0, ymax=1.2,
	enlarge x limits=0.25,
	symbolic x coords={@5, @10, @15},
	xtick=data,
	ylabel style = {font=\small},
	ytick={0, 0.5, 1.0},]
    \addplot[color=orange] coordinates {
    (@5, 0.103) (@10, 0.088) (@15, 0.078)};
    \addplot coordinates {
    (@5, 0.400) (@10, 0.441) (@15, 0.440)};
    \addplot coordinates {
    (@5, 0.336) (@10, 0.272) (@15, 0.240)};
    \addplot coordinates {
    (@5, 0.391) (@10, 0.416) (@15, 0.407)};
    \addplot coordinates {
    (@5, 0.406) (@10, 0.427) (@15, 0.422)};
    \addplot coordinates {
    (@5, 0.410) (@10, 0.422) (@15, 0.405)};
    \addplot coordinates {
    (@5, 0.418) (@10, 0.355) (@15, 0.319)};

	\end{groupplot}
	
	%legend
	\path (myplot c1r1.north west|-current bounding box.center)--
	coordinate(legendpos)
	(myplot c3r1.north east|-current bounding box.north);
    % \path (top|-current bounding box.north)--
    % coordinate(legendpos)
    % (bot|-current bounding box.north);
% 	\hspace{0.6in}
	\matrix[
	matrix of nodes,
	anchor=south,
	text width= 30em,
	row 1/.style = {nodes={font=\footnotesize}},
	draw,
	inner sep=0.2em,
	draw
	]at([yshift=2.5cm]legendpos)
	{
		\ref{plots:tplot1} Abs Greedy
		\ref{plots:tplot2} Max Entropy
		\ref{plots:tplot3} CRM
		\ref{plots:tplot4} EAR
		\ref{plots:tplot5} SCPR
		\ref{plots:tplot6} KG-CRS(ME)
		\ref{plots:tplot7} KG-CRS(GT)\\};
	
    \end{tikzpicture}
    \caption{Comparative results with different $T$ (the maximum number of rounds). %For SR and APA, the higher bar is better while the shorter bar for AT means more efficient CRS
    %and different recommendation list length $K$.
    }
    \label{fig:resultMaxTurn}
\end{figure}

\begin{figure}[htbp]
    \centering
%	\footnotesize
	\begin{tikzpicture}
	\begin{groupplot}[group style={
		group name=myplot,
		group size= 3 by 3,  horizontal sep=1.1cm}, 
		height=4cm, width=3.5cm,
	ylabel style={yshift=-0.1cm},
	legend style = {font=\small, column sep=-1cm},
	every tick label/.append style={font=\footnotesize}
	]
	
	% SR, Yelp
	\hspace{-0.2in}
	\nextgroupplot[ybar=0.10,
	bar width=0.4em,
	ylabel={SR (Yelp)},
	scaled ticks=false,
	yticklabel style={/pgf/number format/.cd,fixed,precision=3},
	ymin=0, ymax=1.2,
	enlarge x limits=0.25,
	symbolic x coords={@5, @10, @15},
	ylabel style = {font=\small},
	xtick=data,
	ytick={0,0.5,1.0},
	]
	\addplot[color=orange] coordinates {
		(@5,0.173) (@10, 0.270) (@15, 0.270)};
	\addplot coordinates {
		(@5,0.922) (@10, 0.929) (@15, 0.929)};
	\addplot coordinates {
		(@5,0.744) (@10, 0.886)  (@15, 0.886)};
	\addplot coordinates {
		(@5,0.840) (@10, 0.962) (@15, 0.962)};
	\addplot coordinates {
		(@5,0.910) (@10, 0.957)  (@15, 0.957)};
	\addplot coordinates {
		(@5,0.862) (@10, 0.944)  (@15, 0.944)};
	\addplot coordinates {
		(@5,0.911) (@10, 0.973)  (@15, 0.973)};

	% AT, Yelp
	\nextgroupplot[ybar=0.1,
	bar width=0.4em,
	ylabel={AT (Yelp)},
	scaled ticks=false,
	yticklabel style={/pgf/number format/.cd,fixed,precision=3},
	ymin=0, ymax=18,
	enlarge x limits=0.25,
	symbolic x coords={@5,@10, @15},
	xtick=data,
	ylabel style = {font=\small},
	ytick={0, 7.5, 15},]
	\addplot[color=orange] coordinates {
		(@5,13.251) (@10, 12.255) (@15, 11.876)};
	\addplot coordinates {
		(@5,6.058) (@10, 5.679) (@15, 5.582)};
	\addplot coordinates {
		(@5,8.130) (@10, 6.157) (@15, 5.553)};
	\addplot coordinates {
		(@5,6.884) (@10, 4.894) (@15, 4.378)};
	\addplot coordinates {
		(@5,6.404) (@10, 5.619) (@15, 5.423)};
    \addplot coordinates {
		(@5,6.123) (@10, 4.798) (@15, 4.452)};
	\addplot coordinates {
		(@5,5.909) (@10, 4.618) (@15, 4.295)};
		
	% APA, Yelp
	\nextgroupplot[ybar=0.1,
	bar width=0.4em,
	ylabel={APA (Yelp)},
	scaled ticks=false,
	yticklabel style={/pgf/number format/.cd,fixed,precision=3},
	ymin=0, ymax=1.2,
	enlarge x limits=0.25,
	symbolic x coords={@5, @10, @15},
	xtick=data,
	ylabel style = {font=\small},
	ytick={0, 0.5, 1.0},]
    \addplot[color=orange] coordinates {
    (@5, 0.013) (@10, 0.022) (@15, 0.023)};
    \addplot coordinates {
    (@5, 0.550) (@10, 0.588) (@15, 0.598)};
    \addplot coordinates {
    (@5, 0.287) (@10, 0.402) (@15, 0.446)};
    \addplot coordinates {
    (@5, 0.388) (@10, 0.571) (@15, 0.638)};
    \addplot coordinates {
    (@5, 0.490) (@10, 0.567) (@15, 0.587)};
    \addplot coordinates {
    (@5, 0.417) (@10, 0.549) (@15, 0.592)};
    \addplot coordinates {
    (@5, 0.508) (@10, 0.663) (@15, 0.713)};
    
	% SR, LastFM
	\nextgroupplot[ybar=0.10,
	bar width=0.4em,
	ylabel={SR (LastFM)},
	scaled ticks=false,
	yticklabel style={/pgf/number format/.cd,fixed,precision=3},
	ymin=0, ymax=1.2,
	enlarge x limits=0.25,
	symbolic x coords={@5, @10, @15},
	ylabel style = {font=\small},
	xtick=data,
	ytick={0,0.5,1.0},
	]
	\addplot[color=orange] coordinates {
		(@5,0.149) (@10, 0.326) (@15, 0.326)};
	\addplot coordinates {
		(@5,0.281) (@10, 0.312) (@15, 0.312)};
	\addplot coordinates {
		(@5,0.104) (@10, 0.275) (@15, 0.275)};
	\addplot coordinates {
		(@5,0.217) (@10, 0.428) (@15, 0.428)};
	\addplot coordinates {
		(@5,0.236) (@10, 0.450)  (@15, 0.450)};
	\addplot coordinates {
		(@5,0.368) (@10, 0.481)  (@15, 0.481)};
	\addplot coordinates {
		(@5,0.363) (@10, 0.512)  (@15, 0.512)};

	% AT, LastFM
	\nextgroupplot[ybar=0.1,
	bar width=0.4em,
	ylabel={AT (LastFM)},
	scaled ticks=false,
	yticklabel style={/pgf/number format/.cd,fixed,precision=3},
	ymin=0, ymax=18,
	enlarge x limits=0.25,
	symbolic x coords={@5,@10, @15},
	xtick=data,
	ylabel style = {font=\small},
	ytick={0, 7.5, 15},]
	\addplot[color=orange] coordinates {
		(@5,13.719) (@10, 12.213) (@15, 11.568)};
	\addplot coordinates {
		(@5,13.784) (@10, 13.419) (@15, 13.328)};
	\addplot coordinates {
		(@5,14.589) (@10, 13.545) (@15, 13.097)};
	\addplot coordinates {
		(@5,13.883) (@10, 12.414) (@15, 11.848)};
	\addplot coordinates {
		(@5,13.970) (@10, 12.732) (@15, 12.255)};
    \addplot coordinates {
		(@5,13.103) (@10, 12.178) (@15, 11.915)};
	\addplot coordinates {
		(@5,13.014) (@10, 11.883) (@15, 11.538)};
		
	% APA, LastFM
	\nextgroupplot[ybar=0.1,
	bar width=0.4em,
	ylabel={APA (LastFM)},
	scaled ticks=false,
	yticklabel style={/pgf/number format/.cd,fixed,precision=3},
	ymin=0, ymax=1.2,
	enlarge x limits=0.25,
	symbolic x coords={@5, @10, @15},
	xtick=data,
	ylabel style = {font=\small},
	ytick={0, 0.5, 1.0},]
    \addplot[color=orange] coordinates {
    (@5, 0.011) (@10, 0.027) (@15, 0.028)};
    \addplot coordinates {
    (@5, 0.219) (@10, 0.228) (@15, 0.229)};
    \addplot coordinates {
    (@5, 0.138) (@10, 0.161) (@15, 0.167)};
    \addplot coordinates {
    (@5, 0.137) (@10, 0.170) (@15, 0.178)};
    \addplot coordinates {
    (@5, 0.183) (@10, 0.218) (@15, 0.226)};
    \addplot coordinates {
    (@5, 0.191) (@10, 0.215) (@15, 0.220)};
    \addplot coordinates {
    (@5, 0.211) (@10, 0.244) (@15, 0.251)};
    
	% SR, Amazon
	\nextgroupplot[ybar=0.10,
	bar width=0.4em,
	ylabel={SR (Amazon)},
	scaled ticks=false,
	yticklabel style={/pgf/number format/.cd,fixed,precision=3},
	ymin=0, ymax=1.2,
	enlarge x limits=0.25,
	symbolic x coords={@5, @10, @15},
	ylabel style = {font=\small},
	xtick=data,
	ytick={0,0.5,1.0},
	]
	\addplot[color=orange] coordinates {
		(@5,0.403) (@10, 0.655) (@15, 0.655)};
	\addplot coordinates {
		(@5,0.775) (@10, 0.835) (@15, 0.835)};
	\addplot coordinates {
		(@5,0.562) (@10, 0.791)  (@15, 0.791)};
	\addplot coordinates {
		(@5,0.744) (@10, 0.892) (@15, 0.892)};
	\addplot coordinates {
		(@5,0.771) (@10, 0.905)  (@15, 0.905)};
	\addplot coordinates {
		(@5,0.742) (@10, 0.906)  (@15, 0.906)};
	\addplot coordinates {
		(@5,0.596) (@10, 0.835)  (@15, 0.835)};

	% AT, Amazon
	\nextgroupplot[ybar=0.1,
	bar width=0.4em,
	ylabel={AT (Amazon)},
	scaled ticks=false,
	yticklabel style={/pgf/number format/.cd,fixed,precision=3},
	ymin=0, ymax=18,
	enlarge x limits=0.25,
	symbolic x coords={@5,@10, @15},
	xtick=data,
	ylabel style = {font=\small},
	ytick={0, 7.5, 15},]
	\addplot[color=orange] coordinates {
		(@5,11.162) (@10, 8.437) (@15, 7.486)};
	\addplot coordinates {
		(@5,10.171) (@10, 9.324) (@15, 9.107)};
	\addplot coordinates {
		(@5,10.223) (@10, 7.739) (@15, 6.882)};
	\addplot coordinates {
		(@5,9.816) (@10, 8.031) (@15, 7.569)};
	\addplot coordinates {
		(@5,9.78) (@10, 7.867) (@15, 7.419)};
    \addplot coordinates {
		(@5,9.75) (@10, 7.861) (@15, 7.351)};
	\addplot coordinates {
		(@5,10.265) (@10, 7.609) (@15, 6.779)};
		
	% APA, Amazon
	\nextgroupplot[ybar=0.1,
	bar width=0.4em,
	ylabel={APA (Amazon)},
	scaled ticks=false,
	yticklabel style={/pgf/number format/.cd,fixed,precision=3},
	ymin=0, ymax=1.2,
	enlarge x limits=0.25,
	symbolic x coords={@5, @10, @15},
	xtick=data,
	ylabel style = {font=\small},
	ytick={0, 0.5, 1.0},]
    \addplot[color=orange] coordinates {
    (@5, 0.036) (@10, 0.078) (@15, 0.088)};
    \addplot coordinates {
    (@5, 0.397) (@10, 0.440) (@15, 0.450)};
    \addplot coordinates {
    (@5, 0.159) (@10, 0.240) (@15, 0.270)};
    \addplot coordinates {
    (@5, 0.318) (@10, 0.407) (@15, 0.432)};
    \addplot coordinates {
    (@5, 0.326) (@10, 0.422) (@15, 0.448)};
    \addplot coordinates {
    (@5, 0.309) (@10, 0.405) (@15, 0.433)};
    \addplot coordinates {
    (@5, 0.213) (@10, 0.319) (@15, 0.357)};
	\end{groupplot}
	
	%legend
	\path (myplot c1r1.north west|-current bounding box.center)--
	coordinate(legendpos)
	(myplot c3r1.north east|-current bounding box.north);
    % \path (top|-current bounding box.north)--
    % coordinate(legendpos)
    % (bot|-current bounding box.north);
% 	\hspace{0.6in}
	\matrix[
	matrix of nodes,
	anchor=south,
	text width= 25em,
	row 1/.style = {nodes={font=\footnotesize}},
	draw,
	inner sep=0.2em,
	draw
	]at([yshift=2.5cm]legendpos)
	{
		\ref{plots:tplot1} Abs Greedy
		\ref{plots:tplot2} Max Entropy
		\ref{plots:tplot3} CRM
		\ref{plots:tplot4} EAR
		\ref{plots:tplot5} SCPR
		\ref{plots:tplot6} KG-CRS(ME)
		\ref{plots:tplot7} KG-CRS(GT)\\};
	
    \end{tikzpicture}
    \caption{Comparative results with different $K$ (the length of recommendation list).% For SR and APA, the higher bar is better while the shorter bar for AT means more efficient CRS
    }%\vspace{-1mm}
    \label{fig:resultK}
\end{figure}
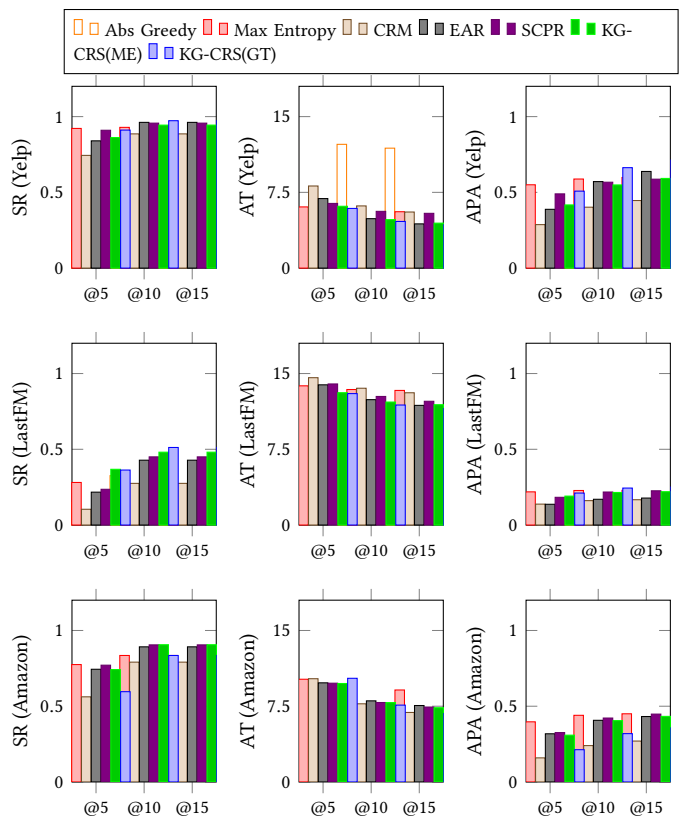

Tables \ref{tb:comparativeResults} and \ref{tb:offlineRec}, and Figures \ref{fig:resultMaxTurn} and \ref{fig:resultK} present the comparative results between KG-CRS and other baselines.

\begin{table*}[t]
\centering
%\small
\caption{Comparative performance of all methods on the three datasets. The best performance is boldfaced, and the runner up is underlined. We compute the improvements that KG-CRS achieves relative to the best baseline. KG-CRS (ME) and KG-CRS (GT) represent our model using max-entropy and ground truth strategies in pre-training, respectively.
Statistical significance of pairwise differences of KG-CRS vs. the best baseline is determined by a paired t-test ($^{***}$ for p-value $\leq$.01, $^{**}$ for p-value $\leq$.05,$^{*}$ for p-value $\leq$.1).\
  }
%\small
%\vspace{-0.1in}
%\addtolength{\tabcep}{4pt}
  \label{tb:comparativeResults}
  \begin{tabular}{l|lll|lll|lll}
    \hline
    &\multicolumn{3}{c|}{Yelp}
    &\multicolumn{3}{c|}{LastFM}
    &\multicolumn{3}{c}{Amazon}\\
    \cline{2-10}
    Method& SR@15 & AT & APA & SR@15 & AT& APA & SR@15 & AT& APA \\
    \hline
    \textbf{Abs Greedy}&0.268&12.239&0.022&0.335&\underline{12.141}&0.028&0.654&8.506&0.077\\
    \textbf{Max Entropy}&0.928&5.695&\underline{0.589}&0.329&13.332&\underline{0.233}&0.836&9.340&\textbf{0.439}\\
    \textbf{CRM}&0.884&6.153&0.408&0.275&13.552&0.160&0.793&\underline{7.669}&0.244\\
    \textbf{EAR}&\underline{0.965}&\underline{4.876}&0.577&0.429&12.435&0.171&0.888&8.024&0.407\\
    \textbf{SCPR}&0.956&5.621&0.571&0.460&12.651&0.222&\textbf{0.907}&7.781&\underline{0.424}\\
    \textbf{KG-CRS (ME)}&0.946&4.925&0.540&\underline{0.477}$^{**}$&12.210&0.216&\underline{0.905}&7.780&0.408\\
    \textbf{KG-CRS (GT)}&\textbf{0.971}&\textbf{4.631}$^{***}$&\textbf{0.664}$^{***}$&\textbf{0.520}$^{***}$&\textbf{11.885}$^{***}$&\textbf{0.245}$^{***}$&0.836&\textbf{7.593}&0.315\\
    \textbf{Improvement}&0.62\%&5.03\%&12.8\%&13.1\%&2.10\%&5.09\%&-&0.98\%&-\\
    \hline
  \end{tabular}
\end{table*}

As depicted in Table \ref{tb:comparativeResults}, we can see that our method KG-CRS performs consistently and largely better than other baselines in terms of the three metrics on Yelp and LastFM, implying the effectiveness of our design for both recommendation and conversation tasks. It should be noted that the comparative results on Amazon are quite mixed in terms of different metrics across different approaches. This might be caused by the larger action space in CRM/EAR/KG-CRS, which leads to difficulty in training the policy network. On the contrary, the action space is fixed as $2$ in SCPR, which makes it perform relatively well on Amazon.

Besides, our method with ground truth strategy, i.e., KG-CRS (GT), is also better than KG-CRS (ME) on Yelp and LastFM, namely the one with max entropy strategy. This confirms the value of ground-truth strategy in pre-training the policy network of RL framework. On Amazon, the results are mixed across KG-CRS (GT) and KG-CRS (ME). By further exploring the wrongly predicted results in test set on Amazon, we find that the poor performance of KG-CRS (GT) is caused by too earlier recommendation, confirming that the relatively worse results of KG-CRS, especially KG-CRS (GT), is caused by the larger action space on Amazon.

\begin{figure*}[h!]
\centering
\includegraphics[scale=0.6]{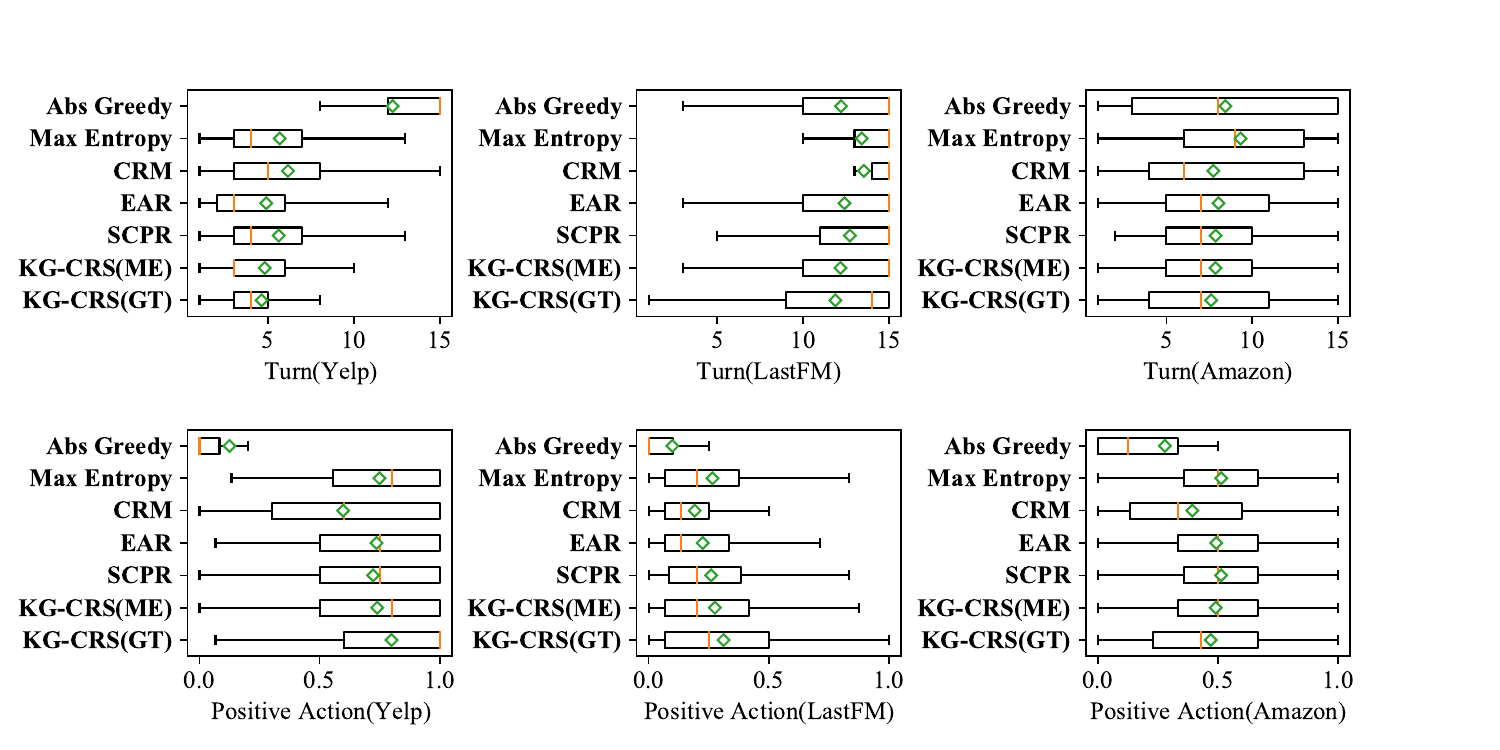}
\caption{Comparative results of all methods in terms of distributions of Turn and Positive Action. The green markers represent the mean of Turn and Positive Action respectively, which are also referred to as AT and APA in Table \ref{tb:comparativeResults}. Each orange line indicates the corresponding median, and the two ends of each rectangle represent the first quartile and the third quartile. Smaller value of Turn shows more effective CRS whilst larger Positive Action indicates better performance.}
\label{fig:distribution}
\end{figure*}

We can obtain similar experimental results by varying $T$ (the maximum number of rounds, see Figure \ref{fig:resultMaxTurn}) and $K$ (the length of recommendation list, see Figure \ref{fig:resultK}). We also report the performance distribution of KG-CRS in terms of AT and APA, as depicted in Figure \ref{fig:distribution}. 

We also compare different approaches in terms of the corresponding recommendation modules. Noted that Abs Greedy and Max Entropy are two methods only designed for the dialogue strategy, while EAR and SCPR use the same recommendation module. Thus, we only compare the recommendation modules of CRM, EAR (SCPR) and KG-CRS.
Table \ref{tb:offlineRec} shows the comparative results of recommendation modules of CRM, EAR (SCPR), and KG-CRS on offline evaluation, i.e., only considering the item ranking task.  Precision, Recall, and NDCG are calculated with the top $10$ items. As can be seen, the design of KG-CRS performs much better than the other two models, further validating its effectiveness on better representation learning for recommendation task in CRS.
\begin{table*}[htb]
\centering
%\footnotesize
\caption{Offline scores of recommendation modules in all methods.
The best performance is boldfaced. }
\label{tb:offlineRec}
    \begin{tabular}{l|l|llll}
        \hline
        Dataset & Method & Precision@10 & Recall@10 & NDCG@10 & AUC\\
        \hline
        & CRM & 0.1068 & 0.6453 & 0.4268 & 0.7987 \\ 
        Yelp & EAR (SCPR) & 0.1089 & 0.6664 & 0.4612 & 0.8321 \\ 
        & KG-CRS & \textbf{0.1142} & \textbf{0.7195} & \textbf{0.5066} & \textbf{0.9057} \\
        \hline
        & CRM & 0.0176 & 0.1047 & 0.0475 & 0.4602 \\ 
        LastFM & EAR (SCPR) & 0.0295 & 0.2234 & 0.1298 & 0.6373 \\ 
        & KG-CRS & \textbf{0.0462} & \textbf{0.3906} & \textbf{0.2370} & \textbf{0.8108} \\
        \hline
        & CRM & 0.0675 & 0.6750 & 0.4106 & 0.6881 \\ 
        Amazon & EAR (SCPR) & 0.0738 & 0.7375 & \textbf{0.4650} & 0.7571 \\ 
        & KG-CRS & \textbf{0.0738} & \textbf{0.7383} & 0.4616 & \textbf{0.7641} \\ 
        \hline
    \end{tabular}
\end{table*}

\subsubsection{Ablation Study (RQ2).} We consider three variants of KG-CRS where `-graphCon', `-graphRec', `-dynamic' denote the model without graph on conversation module, the model without graph on recommendation module, and the model with static graph, respectively. The experimental results are presented in Table \ref{tb:ablationStudy}.
As can be seen, first, under most scenarios, the performance of KG-CRS is better than the other three variants. Specifically, KG-CRS performs almost consistently better than `-graphCon' and `-graphRec', verifying the effectiveness of graph embedding learning techniques for both the recommendation and conversation tasks in CRS. Second, KG-CRS performs slightly worse than `-dynamic' in terms of SR and AT on LastFM.

\begin{table*}[htbp]
\centering
%\small
% \caption{The ablation study considering three variants.}
\caption{The ablation study considering three variants. Here, KG-CRS refers to the better model between KG-CRS (ME) and KG-CRS (GT) on different dataset; `-graphCon' and `-graphRec' do not use the graph in the conversation module and the recommendation module respectively; `-dynamic' keeps static graph for conversation and recommendation modules.
}
% \vspace{-0.1in}
\label{tb:ablationStudy}
    \begin{tabular}{l|lll|lll|lll}
        \hline
        &\multicolumn{3}{c|}{Yelp}
        &\multicolumn{3}{c|}{LastFM}
        &\multicolumn{3}{c}{Amazon}\\
        \cline{2-10}
        Method& SR@15 & AT & APA & SR@15 & AT& APA & SR@15 & AT& APA \\
        \hline
        -graphCon & 0.956 & 4.809 & 0.629 & 0.521 & 12.095 & 0.244 & 0.893 & 7.830 & 0.383 \\ 
        -graphRec & 0.945 & 4.800 & 0.630 & 0.402 & 12.769 & 0.218 & \textbf{0.906} & 7.913 & 0.397 \\ 
        -dynamic & 0.970 & \textbf{4.607} & 0.657 & \textbf{0.547} & \textbf{11.450} & 0.221 & 0.901 & 7.827 & 0.397 \\ 
        \textbf{KG-CRS} & \textbf{0.971} & 4.631 & \textbf{0.664} & 0.520 & 11.885 & \textbf{0.245} & 0.905 & \textbf{7.780} & \textbf{0.408} \\
        \hline
    \end{tabular}
\end{table*}

We further explore the impact of projection matrices (i.e. $M_{\mathcal{0u}}$, $M_{\mathcal{0v}}$, $M_{\mathcal{1v}}$, $M_{\mathcal{1p}}$ in Equation \ref{eq:item_score}, $M_{\mathcal{2u}}$, $M_{\mathcal{2p}}$ in Equation \ref{eq:user-based-prob}, and $M_{\mathcal{1p}}$ in Equation \ref{eq:conv-based-prob}) on learning entity representations.
As shown in Table \ref{tb:albaRec}, on the offline recommendation, KG-CRS performs better than `-map', namely, our model without projection matrices after propagation, demonstrating the necessity of this consideration. Similar conclusions can be drawn from the experimental results on online test, as shown in Table \ref{tb:AlbaMapOnline}.

\begin{table*}[htb]
\centering
%\small
% \caption{The ablation study of the projection matrices on offline recommendation dataset in terms of AUC.}
\caption{Offline scores of recommendation module of ous KG-CRS with or without projection matrices. `-map` represents our model without projection matrices after propagation. 
%Precision, Recall, and NDCG are calculated with the top 10 items. 
The best performance is boldfaced.}
 % \vspace{-0.1in}
\label{tb:albaRec}
    \begin{tabular}{l|l|llll}
        \hline
        Dataset & Method & Precision@10 & Recall@10 & NDCG@10 & AUC\\
        \hline
        & - map & 0.1119 & 0.6961 & 0.4876 & 0.8900 \\ 
        Yelp & \textbf{KG-CRS} & \textbf{0.1142} & \textbf{0.7195} & \textbf{0.5066} & \textbf{0.9057} \\ 
        \hline
        & - map & 0.0429 & 0.3570 & 0.2034 & 0.7892 \\
        LastFM & \textbf{KG-CRS} & \textbf{0.0462} & \textbf{0.3906} & \textbf{0.2370} & \textbf{0.8108} \\
        \hline
        & - map & 0.0733 & 0.7328 & 0.4535 & 0.7596 \\ 
        Amazon & \textbf{KG-CRS} & \textbf{0.0738} & \textbf{0.7383} & \textbf{0.4616} & \textbf{0.7641} \\ 
        \hline
    \end{tabular}
\end{table*}

\begin{table*}[htbp]
\centering
%\small
\caption{Performance of our KG-CRS with or without projection matrices.
%Here KG-CRS refers to the better-performing model between KG-CRS (ME) and KG-CRS (GT) on different dataset. 
The best performance is boldfaced.}
% \vspace{-0.1in}
\label{tb:AlbaMapOnline}
\begin{tabular}{l|lll|lll|lll}
        \hline
        &\multicolumn{3}{c|}{Yelp}
        &\multicolumn{3}{c|}{LastFM}
        &\multicolumn{3}{c}{Amazon}\\
        \cline{2-10}
        Method& SR@15 & AT & APA & SR@15 & AT& APA & SR@15 & AT& APA \\
        \hline
        -map & \textbf{0.974} & \textbf{4.572} & \textbf{0.681} & 0.465 & 12.405 & 0.239 & 0.893 & 8.054 & 0.400 \\ 
        \textbf{KG-CRS} & 0.971 & 4.631 & 0.664 & \textbf{0.520} & \textbf{11.885} & \textbf{0.245} & \textbf{0.905} & \textbf{7.780} & \textbf{0.408} \\
        \hline
    \end{tabular}
\end{table*}

Finally, we also compare the performance between the coarse-grained reward framework and the fine-grained reward one. The experimental results are summarized in Table \ref{tb:fine-grained-reward}. As can be seen, the performance of the two reward frameworks is mixed across different datasets, validating that it is worthwhile to further explore more effective reward designs for CRS.

\begin{table*}[htbp]
\centering
%\small
\caption{Performance of our KG-CRS trained with different reward design. FG reward and CG reward denote fine-grained and coarse-grained reward framework, respectively.}
% \vspace{-0.1in}
\label{tb:fine-grained-reward}
\begin{tabular}{l|lll|lll|lll}
        \hline
        &\multicolumn{3}{c|}{Yelp}
        &\multicolumn{3}{c|}{LastFM}
        &\multicolumn{3}{c}{Amazon}\\
        \cline{2-10}
        Method& SR@15 & AT & APA & SR@15 & AT& APA & SR@15 & AT& APA \\
        \hline
        FG reward & \textbf{0.973} & 4.633 & 0.638 & 0.511 & 11.919 & 0.241 & \textbf{0.918} & \textbf{7.771} & \textbf{0.413} \\ 
        CG reward & 0.971 & \textbf{4.631} & \textbf{0.664} & \textbf{0.520} & \textbf{11.885} & \textbf{0.245} & 0.905 & 7.780 & 0.408 \\
        \hline
    \end{tabular}
\end{table*}

\subsubsection{Sensitivity of Hyper-parameters (RQ3).}
\begin{figure}[htb]
%\footnotesize
\pgfplotsset{every tick label/.append style={font=\small}}
\begin{tikzpicture}
  \begin{axis}[name=plot1,height=4cm,width=3.5cm,x label style={at={(axis description cs:.5,-.1)},anchor=north},y label style={at={(axis description cs:-0.2,.5)},anchor=south}, ylabel=\small{SR$@15$}]
      \addplot [mark=diamond*,mark size=1pt,thick,blue] coordinates {(1,0.496)(2,0.535)(3,0.541)(4,0.559)}; \label{plot:lastFM}
      \addplot [mark=*,mark size=1pt,thick,red] coordinates {(1,0.978)(2,0.973)(3,0.983)(4,0.980)}; \label{plot:Yelp}
      \addplot [mark=square*,mark size=1pt,thick,brown] coordinates {(1,0.8935)(2,0.9095)(3,0.9125)(4,0.906)}; \label{plot:Amazon}
  \end{axis}
  
  \begin{axis}[name=plot2,at={($(plot1.east)+(0.9cm,0)$)},anchor=west,height=4cm,width=3.5cm,x label style={at={(axis description cs:.5,-.1)},anchor=north},y label style={at={(axis description cs:-0.2,.5)},anchor=south},ylabel=\small{AT},legend style={at={($(plot1.east)+(-4cm,1.5cm)$)},anchor=west,legend columns=-1}]
      \addplot [mark=diamond*,mark size=1pt,thick,blue] coordinates {(1,12.320)(2,11.829)(3,11.842)(4,11.640)};
        \addplot [mark=*,mark size=1pt,thick,red] coordinates {(1,4.690)(2,4.562)(3,4.470)(4,4.516)};
        \addplot [mark=square*,mark size=1pt,thick,brown] coordinates {(1,8.175)(2, 7.8685)(3,7.744)(4,7.8605)};
        \legend{\small{lastFM}, \small{Yelp}, \small{Amazon}}
  \end{axis}
  
  \begin{axis}[name=plot3,at={($(plot2.east)+(0.9cm,0cm)$)},anchor=west,height=4cm,width=3.5cm,x label style={at={(axis description cs:.5,-.1)},anchor=north},y label style={at={(axis description cs:-0.2,.5)},anchor=south},ylabel=\small{APA}]
  \addplot [mark=diamond*,mark size=1pt,thick,blue] coordinates {(1,0.244)(2,0.247)(3,0.246)(4,0.250)};
  \addplot [mark=*,mark size=1pt,thick,red] coordinates {(1,0.670)(2,0.670)(3,0.681)(4,0.681)};
  \addplot [mark=square*,mark size=1pt,thick,brown] coordinates {(1,0.393)(2, 0.415)(3,0.411)(4,0.405)};
  \end{axis}
 
\end{tikzpicture}
\caption{Performance of our KG-CRS with different propagation step on the graph.
}
\label{fig:hyperparas}
\end{figure}
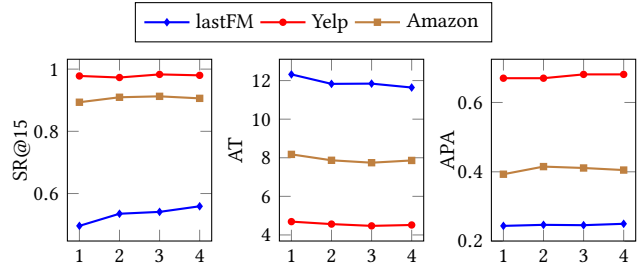

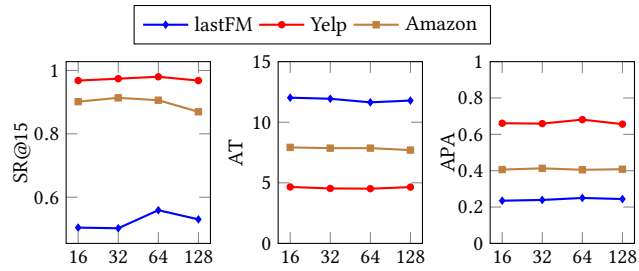
\begin{figure}[h!]
\footnotesize
\pgfplotsset{every tick label/.append style={font=\small}}
\begin{tikzpicture}
  \begin{axis}[name=plot1,height=4cm,width=3.5cm,x label style={at={(axis description cs:.5,-.1)},anchor=north},y label style={at={(axis description cs:-0.2,.5)},anchor=south}, ylabel=\small{SR$@15$}, symbolic x coords={16, 32, 64, 128}]
      \addplot [mark=diamond*,mark size=1pt,thick,blue] coordinates {(16,0.504)(32,0.502)(64,0.559)(128,0.530)}; \label{plot:lastFM}
      \addplot [mark=*,mark size=1pt,thick,red] coordinates {(16,0.968)(32,0.974)(64,0.980)(128,0.968)}; \label{plot:Yelp}
      \addplot [mark=square*,mark size=1pt,thick,brown] coordinates {(16,0.9015)(32,0.9135)(64,0.906)(128,0.8695)}; \label{plot:Amazon}
  \end{axis}
  
  \begin{axis}[name=plot2,at={($(plot1.east)+(0.9cm,0)$)},anchor=west,height=4cm,width=3.5cm,x label style={at={(axis description cs:.5,-.1)},anchor=north},y label style={at={(axis description cs:-0.2,.5)},anchor=south},ylabel=\small{AT},legend style={at={($(plot1.east)+(-4cm,1.7cm)$)},anchor=west,legend columns=-1}, ymin=0, ymax=15, symbolic x coords={16, 32, 64, 128}]
      \addplot [mark=diamond*,mark size=1pt,thick,blue] coordinates {(16,12.023)(32,11.936)(64,11.640)(128,11.787)};
      \addplot [mark=*,mark size=1pt,thick,red] coordinates {(16,4.657)(32,4.536)(64,4.516)(128,4.645)};
      \addplot [mark=square*,mark size=1pt,thick,brown] coordinates {(16,7.9225)(32,7.86)(64,7.8605)(128,7.6995)};
      \legend{\small{lastFM}, \small{Yelp}, \small{Amazon}}
  \end{axis}
  
  \begin{axis}[name=plot3,at={($(plot2.east)+(0.9cm,0cm)$)},anchor=west,height=4cm,width=3.5cm,x label style={at={(axis description cs:.5,-.1)},anchor=north},y label style={at={(axis description cs:-0.2,.5)},anchor=south},ylabel=\small{APA}, ymin=0, ymax=1, symbolic x coords={16, 32, 64, 128}]
  \addplot [mark=diamond*,mark size=1pt,thick,blue] coordinates {(16,0.235)(32,0.239)(64,0.250)(128,0.244)};
  \addplot [mark=*,mark size=1pt,thick,red] coordinates {(16,0.661)(32,0.659)(64,0.681)(128,0.656)};
  \addplot [mark=square*,mark size=1pt,thick,brown] coordinates {(16,0.406)(32,0.413)(64,0.405)(128,0.408)};
  \end{axis}
 
\end{tikzpicture}
\caption{Performance of our KG-CRS with different graph embedding size.}
\label{fig:hyperpara-embed}
\end{figure}

We investigate the impact of the major hyper-parameters, including the propagation step, the embedding size of each step, the four numerical rewards in the policy network, the weight $\beta$ and the discount factor $\gamma$. We apply a grid search in the range of $\{1,2,3,4\}$, and $\{16,32,64,128\}$ to test the impact of the first two hyper-parameters respectively. Weight $\beta$ is set in the range of $\{ 0.01, 0.02, 0.05, 0.07, 0.09, 0.1 \}$ to test its impact. For discount factor $\gamma$, we conduct a grid search in the range of $\{0.5,0.7,0.9,1.0\}$. We vary the value of negative rewards $r_{turn}$ and $r_{quit}$ both in the range of $\{-0.3,-0.15,-0.1,-0.05,-0.01,0\}$, while the value of positive rewards $r_{item}$ and $r_{attr}$ in the range of $\{0, 0.1,0.3,0.5,0.7,1\}$. Figures \ref{fig:hyperparas}-\ref{fig:hyperpara-gamma} show the experimental results of the three hyper-parameters: the propagation step, the embedding size, and $\gamma$, respectively.

\begin{figure}[h!]
\footnotesize
\pgfplotsset{every tick label/.append style={font=\small}}
\begin{tikzpicture}
  \begin{axis}[name=plot1,height=4cm,width=3.5cm,x label style={at={(axis description cs:.5,-.1)},anchor=north},y label style={at={(axis description cs:-0.2,.5)},anchor=south}, ylabel=\small{SR$@15$}, symbolic x coords={0.01, 0.02, 0.05, 0.07, 0.09, 0.1}]
      \addplot [mark=diamond*,mark size=1pt,thick,blue] coordinates {(0.01,0.521)(0.02,0.541)(0.05,0.506)(0.07,0.542)(0.09,0.514)(0.1,0.536)};
      \addplot [mark=*,mark size=1pt,thick,red] coordinates {(0.01,0.952)(0.02,0.973)(0.05,0.936)(0.07,0.918)(0.09,0.957)(0.1,0.981)};
      \addplot [mark=square*,mark size=1pt,thick,brown] coordinates {(0.01,0.9135)(0.02,0.9105)(0.05,0.9175)(0.07,0.918)(0.09,0.914)(0.1,0.9185)};
  \end{axis}
  
  \begin{axis}[name=plot2,at={($(plot1.east)+(0.9cm,0)$)},anchor=west,height=4cm,width=3.5cm,x label style={at={(axis description cs:.5,-.1)},anchor=north},y label style={at={(axis description cs:-0.2,.5)},anchor=south},ylabel=\small{AT},legend style={at={($(plot1.east)+(-4cm,1.7cm)$)},anchor=west,legend columns=-1}, ymin=0, ymax=15, symbolic x coords={0.01, 0.02, 0.05, 0.07, 0.09, 0.1}]
      \addplot [mark=diamond*,mark size=1pt,thick,blue] coordinates {(0.01,11.795)(0.02,11.775)(0.05,11.950)(0.07,11.765)(0.09,11.916)(0.1,11.721)};
      \addplot [mark=*,mark size=1pt,thick,red] coordinates {(0.01,4.883)(0.02,4.766)(0.05,5.099)(0.07,5.275)(0.09,4.846)(0.1,4.584)};
      \addplot [mark=square*,mark size=1pt,thick,brown] coordinates {(0.01,7.7335)(0.02,7.736)(0.05,7.695)(0.07,7.7135)(0.09,7.6595)(0.1,7.7105)}; 
      \legend{\small{lastFM}, \small{Yelp}, \small{Amazon}}
  \end{axis}
  
  \begin{axis}[name=plot3,at={($(plot2.east)+(0.9cm,0cm)$)},anchor=west,height=4cm,width=3.5cm,x label style={at={(axis description cs:.5,-.1)},anchor=north},y label style={at={(axis description cs:-0.2,.5)},anchor=south},ylabel=\small{APA}, ymin=0, ymax=1, symbolic x coords={0.01, 0.02, 0.05, 0.07, 0.09, 0.1}]
  \addplot [mark=diamond*,mark size=1pt,thick,blue] coordinates {(0.01,0.241)(0.02,0.245)(0.05,0.237)(0.07,0.244)(0.09,0.240)(0.1,0.243)};
  \addplot [mark=*,mark size=1pt,thick,red] coordinates {(0.01,0.609)(0.02,0.653)(0.05,0.520)(0.07,0.525)(0.09,0.580)(0.1,0.655)};
  \addplot [mark=square*,mark size=1pt,thick,brown] coordinates {(0.01,0.410)(0.02,0.411)(0.05,0.415)(0.07,0.414)(0.09,0.418)(0.1,0.414)}; 
  \end{axis}
 
\end{tikzpicture}
\caption{Performance of our KG-CRS with different weight $\beta$ (for policy network training).}
\label{fig:hyperpara-beta}
\end{figure}
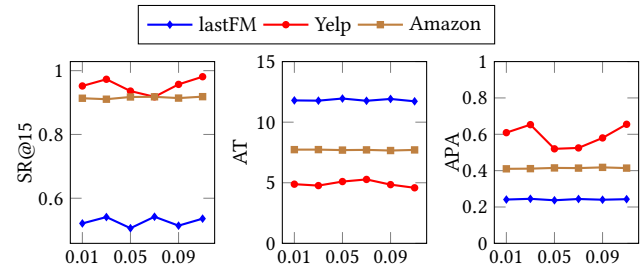

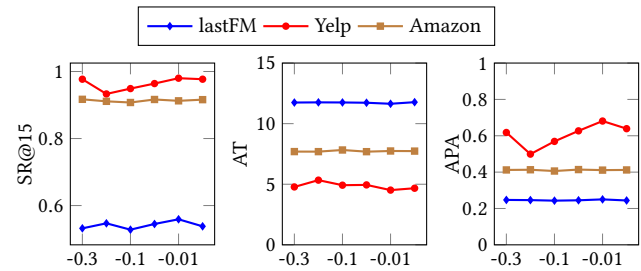
\begin{figure}[h!]
\footnotesize
\pgfplotsset{every tick label/.append style={font=\small}}
\begin{tikzpicture}
  \begin{axis}[name=plot1,height=4cm,width=3.5cm,x label style={at={(axis description cs:.5,-.1)},anchor=north},y label style={at={(axis description cs:-0.2,.5)},anchor=south}, ylabel=\small{SR$@15$}, symbolic x coords={-0.3, -0.15, -0.1, -0.05, -0.01, 0}]
  \addplot [mark=diamond*,mark size=1pt,thick,blue] coordinates {(-0.3,0.532)(-0.15,0.547)(-0.1,0.528)(-0.05,0.545)(-0.01,0.559)(0,0.538)};
  \addplot [mark=*,mark size=1pt,thick,red] coordinates {(-0.3,0.977)(-0.15,0.933)(-0.1,0.949)(-0.05,0.964)(-0.01,0.980)(0,0.977)};
  \addplot [mark=square*,mark size=1pt,thick,brown] coordinates {(-0.3,0.917)(-0.15,0.911)(-0.1,0.9075)(-0.05,0.9165)(-0.01,0.9125)(0,0.916)};
  \end{axis}

  \begin{axis}[name=plot2,at={($(plot1.east)+(0.9cm,0)$)},anchor=west,height=4cm,width=3.5cm,x label style={at={(axis description cs:.5,-.1)},anchor=north},y label style={at={(axis description cs:-0.2,.5)},anchor=south},ylabel=\small{AT},legend style={at={($(plot1.east)+(-4cm,1.7cm)$)},anchor=west,legend columns=-1}, ymin=0, ymax=15, symbolic x coords={-0.3, -0.15, -0.1, -0.05, -0.01, 0}]
      \addplot [mark=diamond*,mark size=1pt,thick,blue] coordinates {(-0.3,11.742)(-0.15,11.756)(-0.1,11.748)(-0.05,11.725)(-0.01,11.640)(0,11.772)};
      \addplot [mark=*,mark size=1pt,thick,red] coordinates {(-0.3,4.777)(-0.15,5.338)(-0.1,4.921)(-0.05,4.945)(-0.01,4.516)(0,4.670)}; 
      \addplot [mark=square*,mark size=1pt,thick,brown] coordinates {(-0.3,7.6935)(-0.15,7.6905)(-0.1,7.8265)(-0.05,7.6875)(-0.01,7.744)(0,7.734)}; 
      \legend{\small{lastFM}, \small{Yelp}, \small{Amazon}}
  \end{axis}

  \begin{axis}[name=plot3,at={($(plot2.east)+(0.9cm,0cm)$)},anchor=west,height=4cm,width=3.5cm,x label style={at={(axis description cs:.5,-.1)},anchor=north},y label style={at={(axis description cs:-0.2,.5)},anchor=south},ylabel=\small{APA}, ymin=0, ymax=1, symbolic x coords={-0.3, -0.15, -0.1, -0.05, -0.01, 0}]
  \addplot [mark=diamond*,mark size=1pt,thick,blue] coordinates {(-0.3,0.247)(-0.15,0.246)(-0.1,0.243)(-0.05,0.245)(-0.01,0.250)(0,0.244)};
  \addplot [mark=*,mark size=1pt,thick,red] coordinates {(-0.3,0.618)(-0.15,0.499)(-0.1,0.569)(-0.05,0.627)(-0.01,0.681)(0,0.639)};
  \addplot [mark=square*,mark size=1pt,thick,brown] coordinates {(-0.3,0.412)(-0.15,0.413)(-0.1,0.406)(-0.05,0.414)(-0.01,0.411)(0,0.412)};
  \end{axis}
 
\end{tikzpicture}
\caption{Performance of our KG-CRS trained with different reward $r_{turn}$.}
\label{fig:hyperpara-r-turn}
\end{figure}

\begin{figure}[h!]
\footnotesize
\pgfplotsset{every tick label/.append style={font=\small}}
\begin{tikzpicture}
  \begin{axis}[name=plot1,height=4cm,width=3.5cm,x label style={at={(axis description cs:.5,-.1)},anchor=north},y label style={at={(axis description cs:-0.2,.5)},anchor=south}, ylabel=\small{SR$@15$}, symbolic x coords={0,0.1,0.3,0.5,0.7,1}]
      \addplot [mark=diamond*,mark size=1pt,thick,blue] coordinates {(0,0.529)(0.1,0.559)(0.3,0.513)(0.5,0.528)(0.7,0.529)(1,0.548)};
      \addplot [mark=*,mark size=1pt,thick,red] coordinates {(0,0.877)(0.1,0.980)(0.3,0.977)(0.5,0.988)(0.7,0.986)(1,0.988)};
      \addplot [mark=square*,mark size=1pt,thick,brown] coordinates {(0,0.909)(0.1,0.9125)(0.3,0.910)(0.5,0.9065)(0.7,0.911)(1,0.9105)};
  \end{axis}
  
  \begin{axis}[name=plot2,at={($(plot1.east)+(0.9cm,0)$)},anchor=west,height=4cm,width=3.5cm,x label style={at={(axis description cs:.5,-.1)},anchor=north},y label style={at={(axis description cs:-0.2,.5)},anchor=south},ylabel=\small{AT},legend style={at={($(plot1.east)+(-4cm,1.7cm)$)},anchor=west,legend columns=-1}, ymin=0, ymax=15, symbolic x coords={0,0.1,0.3,0.5,0.7,1}]
      \addplot [mark=diamond*,mark size=1pt,thick,blue] coordinates {(0,11.849)(0.1,11.640)(0.3,11.978)(0.5,11.834)(0.7,11.720)(1,11.605)};
      \addplot [mark=*,mark size=1pt,thick,red] coordinates {(0,5.720)(0.1,4.516)(0.3,4.759)(0.5,5.014)(0.7,5.091)(1,5.239)};
      \addplot [mark=square*,mark size=1pt,thick,brown] coordinates {(0,7.764)(0.1,7.744)(0.3,7.7695)(0.5,7.7455)(0.7,7.76)(1,7.7815)}; 
      \legend{\small{lastFM}, \small{Yelp}, \small{Amazon}}
  \end{axis}

  \begin{axis}[name=plot3,at={($(plot2.east)+(0.9cm,0cm)$)},anchor=west,height=4cm,width=3.5cm,x label style={at={(axis description cs:.5,-.1)},anchor=north},y label style={at={(axis description cs:-0.2,.5)},anchor=south},ylabel=\small{APA}, ymin=0, ymax=1, symbolic x coords={0,0.1,0.3,0.5,0.7,1}]
      \addplot [mark=diamond*,mark size=1pt,thick,blue] coordinates {(0,0.242)(0.1,0.250)(0.3,0.237)(0.5,0.241)(0.7,0.243)(1,0.251)};
      \addplot [mark=*,mark size=1pt,thick,red] coordinates {(0,0.393)(0.1,0.681)(0.3,0.666)(0.5,0.730)(0.7,0.755)(1,0.767)};
      \addplot [mark=square*,mark size=1pt,thick,brown] coordinates {(0,0.408)(0.1,0.411)(0.3,0.412)(0.5,0.410)(0.7,0.409)(1,0.407)};
  \end{axis}
 
\end{tikzpicture}
\caption{Performance of our KG-CRS trained with different reward $r_{attr}$.}
\label{fig:hyperpara-r-attr}
\end{figure}
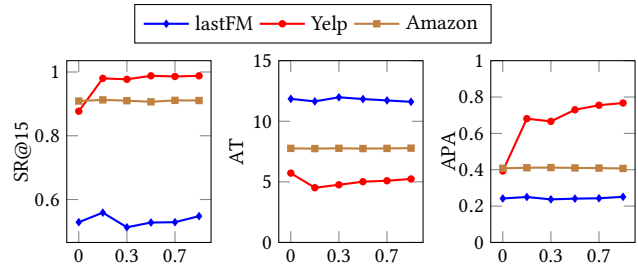

As can be seen, our approach is relatively insensitive to the three hyper-parameters. Specifically, as the propagation step increases (Figure \ref{fig:hyperparas}), KG-CRS performs slightly better in terms of SR@15 and APA on LastFM, while the performance on Yelp, Amazon and AT of LastFM is relatively stable; with the increase of embedding size (Figure \ref{fig:hyperpara-embed}), KG-CRS gets better in terms of SR@15 on LastFM while other metrics are stable; while discount factor increases (Figure \ref{fig:hyperpara-gamma}), the performance of KG-CRS will improve on Yelp only in terms of SR@15 and APA.
For the reward design in the reinforcement learning,
Figures \ref{fig:hyperpara-beta}-\ref{fig:hyperpara-r-attr} demonstrate that KG-CRS is relatively insensitive to the five hyper-parameters, except the $r_{attr}$ on Yelp. This might be caused by that, there are originally $590$ attributes on Yelp. In this case, a larger $r_{attr}$ is preferred if CRS agent identifies a positive attribute.

% gamma
\begin{figure}[h!]
\footnotesize
\pgfplotsset{every tick label/.append style={font=\small}}
\begin{tikzpicture}
  \begin{axis}[name=plot1,height=4cm,width=3.5cm,x label style={at={(axis description cs:.5,-.1)},anchor=north},y label style={at={(axis description cs:-0.2,.5)},anchor=south}, ylabel=\small{SR$@15$}, symbolic x coords={0.5,0.7,0.9, 1.0}]
      \addplot [mark=diamond*,mark size=1pt,thick,blue] coordinates {(0.5,0.533)(0.7,0.559)(0.9,0.529)(1.0,0.540)};
      \addplot [mark=*,mark size=1pt,thick,red] coordinates {(0.5,0.908)(0.7,0.980)(0.9,0.984)(1.0,0.966)};
      \addplot [mark=square*,mark size=1pt,thick,brown] coordinates {(0.5,0.905)(0.7,0.9125)(0.9,0.9155)(1.0,0.915)};
  \end{axis}
  
  \begin{axis}[name=plot2,at={($(plot1.east)+(0.9cm,0)$)},anchor=west,height=4cm,width=3.5cm,x label style={at={(axis description cs:.5,-.1)},anchor=north},y label style={at={(axis description cs:-0.2,.5)},anchor=south},ylabel=\small{AT},legend style={at={($(plot1.east)+(-4cm,1.7cm)$)},anchor=west,legend columns=-1}, ymin=0, ymax=15, symbolic x coords={0.5,0.7,0.9, 1.0}]
      \addplot [mark=diamond*,mark size=1pt,thick,blue] coordinates {(0.5,11.779)(0.7,11.640)(0.9,11.771)(1.0,11.734)};
      \addplot [mark=*,mark size=1pt,thick,red] coordinates {(0.5,5.481)(0.7,4.516)(0.9,4.649)(1.0,5.342)}; 
      \addplot [mark=square*,mark size=1pt,thick,brown] coordinates {(0.5,7.8535)(0.7,7.744)(0.9,7.7125)(1.0,7.7375)};
      \legend{\small{lastFM}, \small{Yelp}, \small{Amazon}}
  \end{axis}
  
  \begin{axis}[name=plot3,at={($(plot2.east)+(0.9cm,0cm)$)},anchor=west,height=4cm,width=3.5cm,x label style={at={(axis description cs:.5,-.1)},anchor=north},y label style={at={(axis description cs:-0.2,.5)},anchor=south},ylabel=\small{APA}, ymin=0, ymax=1, symbolic x coords={0.5,0.7,0.9, 1.0}]
  \addplot [mark=diamond*,mark size=1pt,thick,blue] coordinates {(0.5,0.244)(0.7,0.250)(0.9,0.242)(1.0,0.243)};
  \addplot [mark=*,mark size=1pt,thick,red] coordinates {(0.5,0.443)(0.7,0.681)(0.9,0.662)(1.0,0.680)}; 
  \addplot [mark=square*,mark size=1pt,thick,brown] coordinates {(0.5,0.405)(0.7,0.411)(0.9,0.414)(1.0,0.413)};
  \end{axis}
 
\end{tikzpicture}
\caption{Performance of our KG-CRS with different discount factor $\gamma$ (for policy network training).}
\label{fig:hyperpara-gamma}
\end{figure}
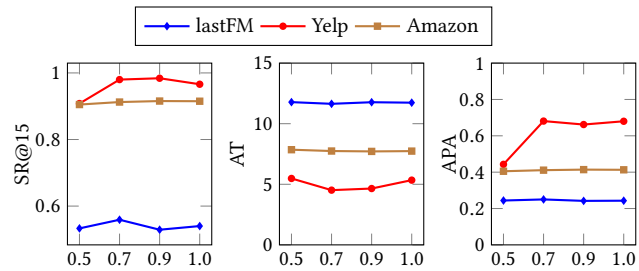

% beta

% r-turn

\newpage
% r-quit
\begin{figure}[h!]
\footnotesize
\pgfplotsset{every tick label/.append style={font=\small}}
\begin{tikzpicture}
  \begin{axis}[name=plot1,height=4cm,width=3.5cm,x label style={at={(axis description cs:.5,-.1)},anchor=north},y label style={at={(axis description cs:-0.2,.5)},anchor=south}, ylabel=\small{SR$@15$}, symbolic x coords={-0.3, -0.15, -0.1, -0.05, -0.01, 0}]
      \addplot [mark=diamond*,mark size=1pt,thick,blue] coordinates {(-0.3,0.559)(-0.15,0.542)(-0.1,0.521)(-0.05,0.547)(-0.01,0.523)(0,0.548)};
      \addplot [mark=*,mark size=1pt,thick,red] coordinates {(-0.3,0.980)(-0.15,0.952)(-0.1,0.931)(-0.05,0.958)(-0.01,0.942)(0,0.962)};
      \addplot [mark=square*,mark size=1pt,thick,brown] coordinates {(-0.3,0.9125)(-0.15,0.915)(-0.1,0.9125)(-0.05,0.9135)(-0.01,0.915)(0,0.9155)};
  \end{axis}

  \begin{axis}[name=plot2,at={($(plot1.east)+(0.9cm,0)$)},anchor=west,height=4cm,width=3.5cm,x label style={at={(axis description cs:.5,-.1)},anchor=north},y label style={at={(axis description cs:-0.2,.5)},anchor=south},ylabel=\small{AT},legend style={at={($(plot1.east)+(-4cm,1.7cm)$)},anchor=west,legend columns=-1}, ymin=0, ymax=15, symbolic x coords={-0.3, -0.15, -0.1, -0.05, -0.01, 0}]
      \addplot [mark=diamond*,mark size=1pt,thick,blue] coordinates {(-0.3,11.640)(-0.15,11.721)(-0.1,11.862)(-0.05,11.689)(-0.01,11.844)(0,11.716)};
      \addplot [mark=*,mark size=1pt,thick,red] coordinates {(-0.3,4.516)(-0.15,4.983)(-0.1,5.321)(-0.05,4.919)(-0.01,4.917)(0,4.729)}; 
      \addplot [mark=square*,mark size=1pt,thick,brown] coordinates {(-0.3,7.744)(-0.15,7.737)(-0.1,7.6995)(-0.05,7.7515)(-0.01,7.7265)(0,7.7135)};
      \legend{\small{lastFM}, \small{Yelp}, \small{Amazon}}
  \end{axis}

  \begin{axis}[name=plot3,at={($(plot2.east)+(0.9cm,0cm)$)},anchor=west,height=4cm,width=3.5cm,x label style={at={(axis description cs:.5,-.1)},anchor=north},y label style={at={(axis description cs:-0.2,.5)},anchor=south},ylabel=\small{APA}, ymin=0, ymax=1, symbolic x coords={-0.3, -0.15, -0.1, -0.05, -0.01, 0}]
  \addplot [mark=diamond*,mark size=1pt,thick,blue] coordinates {(-0.3,0.250)(-0.15,0.242)(-0.1,0.242)(-0.05,0.248)(-0.01,0.240)(0,0.247)};
  \addplot [mark=*,mark size=1pt,thick,red] coordinates {(-0.3,0.681)(-0.15,0.547)(-0.1,0.511)(-0.05,0.598)(-0.01,0.601)(0,0.623)};
  \addplot [mark=square*,mark size=1pt,thick,brown] coordinates {(-0.3,0.411)(-0.15,0.413)(-0.1,0.412)(-0.05,0.413)(-0.01,0.412)(0,0.414)};
  \end{axis}
 
\end{tikzpicture}
\caption{Performance of our KG-CRS trained with different reward $r_{quit}$.}
\label{fig:hyperpara-r-quit}
\end{figure}
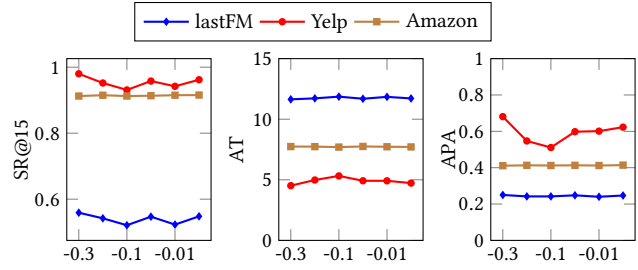

% r-item
\begin{figure}[h!]
\footnotesize
\pgfplotsset{every tick label/.append style={font=\small}}
\begin{tikzpicture}
  \begin{axis}[name=plot1,height=4cm,width=3.5cm,x label style={at={(axis description cs:.5,-.1)},anchor=north},y label style={at={(axis description cs:-0.2,.5)},anchor=south}, ylabel=\small{SR$@15$}, symbolic x coords={0,0.1,0.3,0.5,0.7,1}]
      \addplot [mark=diamond*,mark size=1pt,thick,blue] coordinates {(0,0.536)(0.1,0.529)(0.3,0.547)(0.5,0.549)(0.7,0.538)(1,0.559)};
      \addplot [mark=*,mark size=1pt,thick,red] coordinates {(0,0.965)(0.1,0.986)(0.3,0.988)(0.5,0.979)(0.7,0.976)(1,0.980)};
      \addplot [mark=square*,mark size=1pt,thick,brown] coordinates {(0,0.914)(0.1,0.913)(0.3,0.9075)(0.5,0.9075)(0.7,0.918)(1,0.9125)};
  \end{axis}
  
  \begin{axis}[name=plot2,at={($(plot1.east)+(0.9cm,0)$)},anchor=west,height=4cm,width=3.5cm,x label style={at={(axis description cs:.5,-.1)},anchor=north},y label style={at={(axis description cs:-0.2,.5)},anchor=south},ylabel=\small{AT},legend style={at={($(plot1.east)+(-4cm,1.7cm)$)},anchor=west,legend columns=-1}, ymin=0, ymax=15, symbolic x coords={0,0.1,0.3,0.5,0.7,1}]
      \addplot [mark=diamond*,mark size=1pt,thick,blue] coordinates {(0,11.775)(0.1,11.812)(0.3,11.752)(0.5,11.585)(0.7,11.750)(1,11.640)};
      \addplot [mark=*,mark size=1pt,thick,red] coordinates {(0,6.229)(0.1,5.483)(0.3,4.997)(0.5,4.964)(0.7,4.837)(1,4.516)}; 
      \addplot [mark=square*,mark size=1pt,thick,brown] coordinates {(0,7.759)(0.1,7.7275)(0.3,7.8255)(0.5,7.7465)(0.7,7.714)(1,7.744)}; 
      \legend{\small{lastFM}, \small{Yelp}, \small{Amazon}}
  \end{axis}

  \begin{axis}[name=plot3,at={($(plot2.east)+(0.9cm,0cm)$)},anchor=west,height=4cm,width=3.5cm,x label style={at={(axis description cs:.5,-.1)},anchor=north},y label style={at={(axis description cs:-0.2,.5)},anchor=south},ylabel=\small{APA}, ymin=0, ymax=1, symbolic x coords={0,0.1,0.3,0.5,0.7,1}]
      \addplot [mark=diamond*,mark size=1pt,thick,blue] coordinates {(0,0.245)(0.1,0.244)(0.3,0.246)(0.5,0.252)(0.7,0.246)(1,0.250)};
      \addplot [mark=*,mark size=1pt,thick,red] coordinates {(0,0.668)(0.1,0.715)(0.3,0.727)(0.5,0.683)(0.7,0.622)(1,0.681)}; 
      \addplot [mark=square*,mark size=1pt,thick,brown] coordinates {(0,0.412)(0.1,0.408)(0.3,0.406)(0.5,0.409)(0.7,0.413)(1,0.411)};
  \end{axis}
 
\end{tikzpicture}
\caption{Performance of our KG-CRS trained with different reward $r_{item}$.}
\label{fig:hyperpara-r-item}
\end{figure}
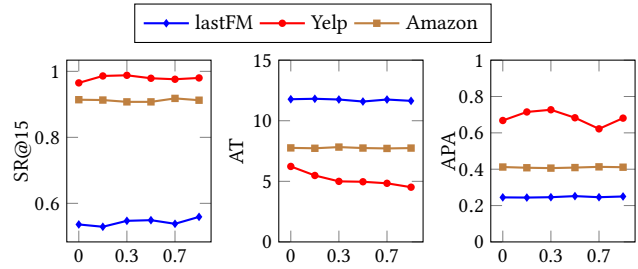

% r-attr

\section{Conclusions}
\label{sec:conclusions}
In this paper, we designed a knowledge-graph based CRS framework particularly for multi-round recommendation scenario to simultaneously improve the recommendation and conversation tasks.
Specifically, we combined user-item graph (built on historical interactions) and item-attribute graph, and dynamically updated the graph by removing negative items and attributes identified by previous rounds in a session. We adopted the graph embedding learning technique on the dynamic graph to better model the interactive relationships among users, items and attributes, where the learned corresponding embedding are integrated into the recommendation module and conversational module.
Extensive experiments on three datasets validated that KG-CRS can perform better than the state-of-the-art approaches in terms of both recommendation and conversation tasks in both online and offline settings. Besides, the ablation study verified the effectiveness of our designs.

For future work, we will consider to design more effective rewards and state vector in the policy network to further improve the performance of conversational recommendation systems. Besides, we will consider to design user studies to empirically verify the effectiveness of CRS and thus design more realistic evaluation metrics towards CRS.

%\bibliographystyle{ACM-Reference-Format}
%\bibliography{paper}


\section{Acknowledgements}
We greatly acknowledge the support of the National Natural Science Foundation of China (Grant No. 72371148), and the Shanghai Rising-Star Program (Grant No. 23QA1403100).

\end{document}